\documentclass[a4paper,aps,prd,10pt,preprintnumbers,showpacs,twocolumn,superscriptaddress,nofootinbib,amsmath,amssymb]{revtex4-1}
\usepackage{graphicx}
\usepackage{cmap}
\usepackage[utf8]{inputenc}
\usepackage[T1]{fontenc} 
\def\imo{i}

\def\K{{\cal K}}
\def\Order#1{{\cal O}\left(#1\right)}

\begin{document}
\title{Analytical QNMs of fields of various spin in the Hayward spacetime}
\author{Zainab Malik}\affiliation{Institute of Applied Sciences and Intelligent Systems, H-15, Islamabad, Pakistan}\email{zainabmalik8115@gmail.com}
\begin{abstract}
By employing an expansion in terms of the inverse multipole number, we derive analytic expressions for the quasinormal modes (QNMs) of scalar, Dirac, and Maxwell perturbations in the Hayward black hole (BH) background. The metric has three interpretations: as a model for a radiating BH, as a quantum-corrected BH owing to the running gravitational coupling in the Asymptotically Safe Gravity, and as a BH solution in the Effective Field Theory. We show that the obtained compact analytical formulas approximate QNMs with remarkable accuracy for $\ell > 0$.
\end{abstract}
\maketitle

\textbf{Introduction.} The proper oscillations (quasinormal) frequencies are observed with the help of gravitational interferometers and telescopes via observations of gravitational and electromagnetic radiation \cite{LIGOScientific:2016aoc,EventHorizonTelescope:2019dse,Goddi:2016qax} and the near future experiments should broaden the observational frequency range \cite{Babak:2017tow} with a potential to constrain the BH's parameters  \cite{Tsukamoto:2014tja,Shaikh:2021yux}. However, considerable uncertainty with which the mass and spin of the just formed BH is measured makes it possible to ascribe signals to a non-Kerr spacetimes, which justify consideration of various alternative/modified theories of gravity.

Singularity appearing in the Schwarzschild solution of the Einstein equations demonstrates the limits of General Relativity. Various alternative theories of gravity, such as asymptotically safe gravity, higher curvature corrected theories, non-linear electrodynamics, and various phenomenological-like approaches, have been considered in the literature in order to solve the above problem. 

Appearance of the regular BH metrics in various approaches to quantum gravity induced a vast literature on perturbations and QNMs of such BHs \cite{Bronnikov:2019sbx,Panotopoulos:2019qjk,Yang:2021cvh,Wahlang:2017zvk,Saleh:2018hba,Toshmatov:2015wga,Wu:2022eiv,Xi:2016qrg,Jusufi:2020odz,Guo:2024jhg,Franzin:2022iai,Al-Badawi:2024mco}.
One of the interesting branches in this direction is connected to the Hayward BHs \cite{Hayward:2005gi},  the quasinormal spectrum of which has been investigated in \cite{DuttaRoy:2022ytr,Al-Badawi:2023lke,Pedraza:2021hzw,Lin:2013ofa}.  The Hayward metric describes the formation of a locally defined BH from an initial vacuum region.

The Hayward solution also appears, up to the redefinition of parameters, in the Effective Field Theory \cite{Mukohyama:2023xyf,Konoplya:2023ppx} and Asymptotically Safe Gravity \cite{Held:2019xde,Konoplya:2023ppx}.
Thus, there are overall at least three interpretations of the Hayward metric, and the study of its spectrum is interesting for applications in various models of gravity. At the same time, all the above studies of quasinormal spectra of regular BHs, including the Hayward one, were done numerically, sometimes providing only an estimate for the eikonal limit. Here, we are interested in the analytical study of QNMs of the Hayward BHs for scalar, Dirac, and electromagnetic fields.

The search for analytic formulas, even if approximate,  for the quasinormal spectrum has a number of well-grounded motivations. In most cases quasinormal frequencies of BHs could be obtained only in a numeric form, because the corresponding wave equation (or, frequently, system of equations) is too complex. There are only several exceptions: three-dimensional Bañados-Teitelboim-Zanell-like spacetimes \cite{Banados:1992wn}. However, even  the generalization of the Bañados-Teitelboim-Zanell spacetimes to include higher curavture terms \cite{Konoplya:2020ibi}  require exploring numerical approaches \cite{Skvortsova:2023zmj}, because the metric function is too complex to allow for the exact analytic soltuion.
Some other exceptions include approximate analytical expressions in the near extreme asymptotically de Sitter spacetime \cite{Cardoso:2003sw,Molina:2003ff} ro the limit of the infinite multipole number $\ell$. At the same time, even approximate analytical expression for QNMs allows us to know the explicit dependence of the frequencies on the BH's parameters and then use a number of tools which are applicable to analytical expressions.
Despite the eikonal formula at $\ell \rightarrow \infty$ cannot be applied to first several multipoles, because it is inaccurate at small $\ell$, expansion beyond the eikonal limit at a few orders produces remarkably precise expressions as  shown in \cite{Konoplya:2023moy}.
This approach has been recently used to find the analytical expressions for QNMs of the Reissner-Nordström-like BHs \cite{doi:10.1142/S0217751X24500246} and holonomy corrected BHs \cite{Bolokhov:2023bwm}.

Within this framework, it is worth mentioning that high multipole numbers, corresponding to high real oscillation frequencies, are interesting to study for a number of reasons.
First of all,  a correspondence between the eikonal QNMs and unstable null geodesics was established in \cite{Cardoso:2008bp}. This correspondence  connects the real and imaginary parts of the eikonal QNMs with the rotational frequency and Lyapunov exponent, respectively, taken at the unstable circular null geodesic. Although this correspondence is valid in the majority of cases, there are exceptions and limits described in \cite{Khanna:2016yow,Konoplya:2019hml,Bolokhov:2023dxq,Konoplya:2017wot,Konoplya:2022gjp}. Another interesting feature of the eikonal regime is that it may bring about the so-called eikonal instability \cite{Takahashi:2011du,Dotti:2004sh,Gleiser:2005ra,Cuyubamba:2016cug,Konoplya:2017lhs}. It is called eikonal because, counterintuitively, the lower multipoles are stable, while the higher ones bring about instability developing at late times, after a stage of damped oscillations \cite{Konoplya:2008ix}. Keeping the above motivations in mind, an eikonal formula for QNMs was derived for perturbations for various BH models: for various Einstein-scalar-Gauss-Bonnet theories \cite{Konoplya:2019hml,Paul:2023eep,Konoplya:2020bxa}, for various quantum corrected theories  \cite{Bolokhov:2023bwm,Bolokhov:2023dxq}, for the Schwarzschild-de Sitter spacetime  \cite{Zhidenko:2003wq}, for the dynamical Chern-Simons theory  \cite{Chen:2022nlw}, for the scalar-tensor theories  \cite{Silva:2019scu}, for magnetically charged BHs  \cite{Toshmatov:2019gxg}, and for BHs with quadrupole momentum  \cite{Allahyari:2018cmg}.



\textbf{Basic equations: metric, wave-like equations and effective potentials.} The background (unperturbed) metric of the Hayward BH is,
\begin{equation}\label{metric}
  ds^2=-f(r)dt^2+ f^{-1}(r) dr^2+r^2(d\theta^2+\sin^2\theta d\phi^2),
\end{equation}
where,
$f(r)=\displaystyle 1- 2 r^2 M^{-2}(\gamma +r^3 M^{-3})^{-1}$ \cite{Berglund:2012bu}.
Here $\gamma $ is the coupling parameter, and $M$ is the mass parameter. Here we  measure all dimensional quantities in units of $M$, i.~e., we take $M=1$.
The event horizon exists if $ \gamma < 32/27 \approx 1.18$, so that in our data we will use the range  $0< \gamma < 1.18$. From the form of the metric function $f(r)$ one can see that in the limit $r \rightarrow 0$ the metric is regular, unlike the Schwarzschild spacetime.

\begin{figure}
\resizebox{\linewidth}{!}{\includegraphics{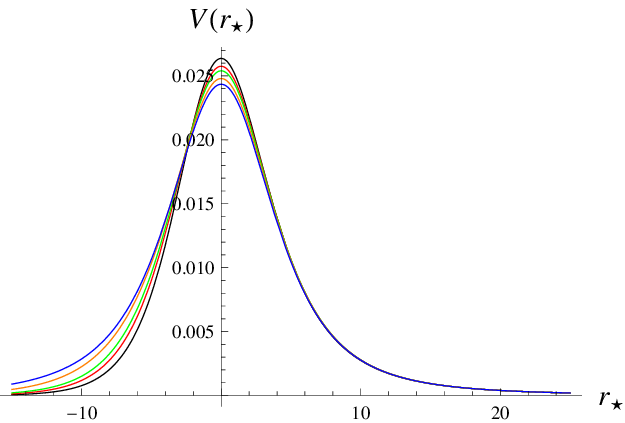}\includegraphics{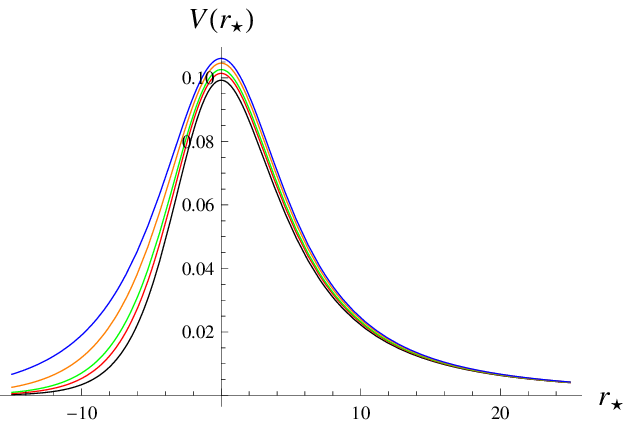}}
\caption{Effective potentials as functions of $r_{*}$ of the $\ell=0$ (left) and $\ell=1$ (right) scalar field for the Hayward spacetime: $\gamma=0$ (black) $\gamma=0.4$ (red) $\gamma=0.6$ (green) $\gamma=0.9$ (orange), and $\gamma=1.1$ (blue); $M=1$.}\label{fig:potentials1}
\end{figure}

\begin{figure}
\resizebox{\linewidth}{!}{\includegraphics{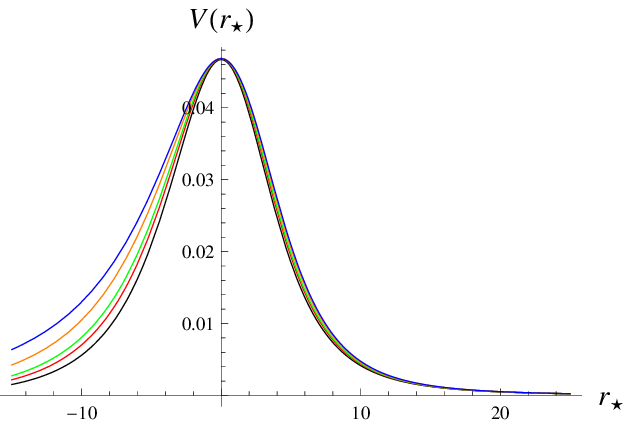}\includegraphics{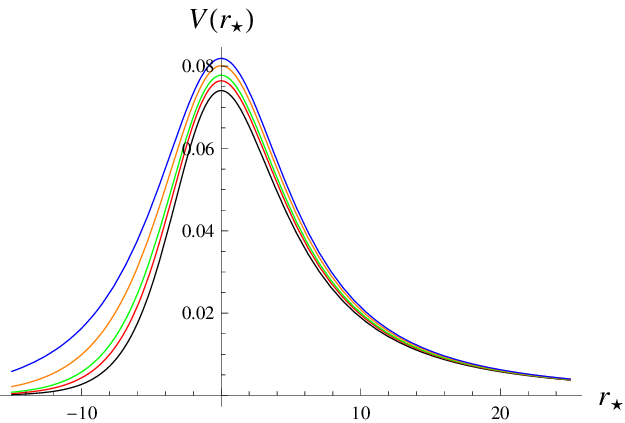}}
\caption{Effective potentials as functions of $r_{*}$ of the $\ell=1/2$ Dirac (left) and $\ell=1$ Maxwell fields for the Hayward spacetime: $\gamma=0$ (black) $\gamma=0.4$ (red) $\gamma=0.6$ (green) $\gamma=0.9$ (orange), and $\gamma=1.1$ (blue); $M=1$.}\label{fig:potentials3}
\end{figure}

The general relativistic scalar, Maxwell, and Dirac fields obey the following covariant equations:
\begin{subequations}\label{coveqs}
\begin{eqnarray}\label{KGg}
\frac{1}{\sqrt{-g}}\partial_\mu \left(\sqrt{-g}g^{\mu \nu}\partial_\nu \Phi\right)&=&0, \quad scalar
\\\label{EmagEq}
\frac{1}{\sqrt{-g}}\partial_{\mu} \left(F_{\rho\sigma}g^{\rho \nu}g^{\sigma \mu}\sqrt{-g}\right)&=&0\,, \quad Maxwell
\\\label{covdirac}
\gamma^{\alpha} \left( \frac{\partial}{\partial x^{\alpha}} - \Gamma_{\alpha} \right) \Psi_{Dirac}&=&0, \quad Dirac
\end{eqnarray}
\end{subequations}
where $F_{\mu\nu}=\partial_\mu A_\nu-\partial_\nu A_\mu$ is the electromagnetic tensor, $\gamma^{\alpha}$ are (noncommutative) gamma matrices and $\Gamma_{\alpha}$ are spin connections.
After separation of variables the above covariant equations (\ref{coveqs}) are transformed to the wavelike form with an effective potential (see, for instance, reviews \cite{Kokkotas:1999bd,Konoplya:2011qq}):
\begin{equation}\label{wave-equation}
\dfrac{d^2 \Psi}{dr_*^2}+(\omega^2-V(r))\Psi=0,
\end{equation}
where to move the event horizon to $-\infty$ we introduce the new radial variable $r_*$, $dr_*\equiv dr/f(r)$.
The corresponding effective potentials for the scalar ($s=0$) and Maxwell ($s=1$) fields have the form
\begin{equation}\label{potentialScalar}
V(r)=f(r)\frac{\ell(\ell+1)}{r^2}+\frac{1-s}{r}\cdot\frac{d^2 r}{dr_*^2},
\end{equation}
where $\ell=s, s+1, s+2, \ldots$ are the multipole numbers.
For the Dirac field ($s=\pm 1/2$) allowing for the two chiralities one has two potentials, which produce the same spectrum
\begin{equation}
V_{\pm}(r) = W^2\pm\frac{dW}{dr_*}, \quad W\equiv \left(\ell+\frac{1}{2}\right)\frac{\sqrt{f(r)}}{r}.
\end{equation}
This happens because the wave functions can be transformed one into another via the well-known Darboux transformation,
\begin{equation}\label{psi}
\Psi_{+}\propto \left(W+\dfrac{d}{dr_*}\right) \Psi_{-}.
\end{equation}
Therefore, we study  QNMs for only one of the two isospectral effective potentials, $V_{+}(r)$. This choice provides better accuracy of the WKB method.
At $\ell >0$ the effective potentials under consideration are positive definite and have single maximum, as shown in figs. \ref{fig:potentials1} and \ref{fig:potentials3}.
The positive definite effective potential guarantees the stability of the perturbation and, thereby, absence of the growing QNMs in the spectrum. 

\textbf{WKB approach.} QNMs satisfy the following boundary conditions,
\begin{equation}\label{boundaryconditions}
\Psi(r_*\to\pm\infty)\propto e^{\pm\imo \omega r_*}.
\end{equation}
These boundary conditions correspond to purely ingoing waves at the event horizon and purely outgoing wave at spatial infinity. In other words, no waves can come from either the horizon or the far zone, because the frequencies are proper and measured when the perturbations source does not act anymore.
\par

If the effective potential $V(r)$ has a barrier shape with one maximum and slowly decaying at the extremities $r_{*}=\pm \infty$, then the WKB formula developed first by Will and Schutz \cite{Schutz:1985km} can be used for finding the low lying QNMs.
The WKB method is based on the matching of the WKB solutions at the extremities with the Taylor series around the potential's maximum. At the first order, the WKB formula is exact when $\ell \to \infty$. Then, at a given higher order the WKB formula for the frequencies can be written as expansions around this, exact, limit \cite{Konoplya:2019hlu}:
\begin{eqnarray}\label{WKBformula-spherical}
\omega^2&=&V_0+A_2(\K^2)+A_4(\K^2)+A_6(\K^2)+\ldots\\\nonumber&-&\imo \K\sqrt{-2V_2}\left(1+A_3(\K^2)+A_5(\K^2)+A_7(\K^2)\ldots\right).
\end{eqnarray}
The limit $\ell \to \infty$ leads to infinite frequency of oscillations and corresponds to the regime of geometrical optics, i.e. the eikonal. Then, the matching conditions produced when the QNMs boundary conditions are imposed,  lead to the following relation \cite{Konoplya:2019hlu}
\begin{equation}
\K=n+\frac{1}{2}, \quad n=0,1,2,\ldots,
\end{equation}
where $n$ is overtone number, $V_0$ is the potential's value in the maximum, while $V_i$ is the i-th derivative of the potential respectively $r_{*}$,  the functions $A_i$ for $i=2, 3, 4, \ldots$ represent $i-th$ order correction terms beyond the eikonal limit. These correction terms depend on $\K$ and derivatives of the effective potential up to the order $2i$.

The explicit form of $A_i$ are written down in \cite{Iyer:1986np} up to the third WKB order, in \cite{Konoplya:2003ii} up to the 6th order and in \cite{Matyjasek:2017psv} up to the 13th order. The WKB approach for finding of quasinormal frequencies and transmission coefficients have been broadly used, so that here we are able to mention only some of these works where further details could be found \cite{Konoplya:2001ji,Konoplya:2006ar,Konoplya:2018ala,Kodama:2009bf,Gong:2023ghh,Skvortsova:2023zmj,Fernando:2012yw}. 
To verify the analytic formula, we employed a more accurate version of the 6th-order WKB method with Padé approximants \cite{Matyjasek:2017psv}. The quantity $\tilde{m}$, which determines the structure of the Padé approximants, is defined in Eq. 21 of \cite{Konoplya:2019hlu}. In this work, we use $\tilde{m}=4$, as it generally provides the best accuracy in the majority of cases (see, for instance, \cite{Skvortsova:2023zmj,Bolokhov:2023dxq,Dubinsky:2024hmn}). However, the choice $\tilde{m}=5$ \cite{Bolokhov:2023bwm,Konoplya:2020jgt} usually also provides very similar results.
\begin{table*}
\begin{tabular}{c c c c c}
\hline
\hline
$\gamma$ & $\widetilde{m}=4$ & analytic & r.e. $Re (\omega)$ & r.e. $Im (\omega)$  \\
\hline
$0$ & $0.110792-0.104683 i$ & $0.108699-0.109788 i$ & $1.89\%$ & $4.88\%$\\
$0.02$ & $0.110951-0.104482 i$ & $0.108824-0.109401 i$ & $1.92\%$ & $4.71\%$\\
$0.04$ & $0.111108-0.104278 i$ & $0.108948-0.109012 i$ & $1.94\%$ & $4.54\%$\\
$0.06$ & $0.111261-0.104071 i$ & $0.109071-0.108618 i$ & $1.97\%$ & $4.37\%$\\
$0.08$ & $0.111410-0.103865 i$ & $0.109194-0.108221 i$ & $1.99\%$ & $4.19\%$\\
$0.1$ & $0.111554-0.103660 i$ & $0.109315-0.107821 i$ & $2.01\%$ & $4.01\%$\\
$1.18$ & $0.110839-0.088116 i$ & $0.112852-0.078891 i$ & $1.82\%$ & $10.5\%$\\
\hline
\hline
\end{tabular}
\caption{QNMs for the $\ell=0$ ($n=0$) scalar field for the Hayward BH obtained by the analytic formula and the 6th order WKB formula with Padé approximants for various  $\gamma$ and $M=1$.}
\end{table*}

\begin{table*}
\centering
\resizebox{\textwidth}{!}{
\begin{tabular}{c c c c c c c c c c}
\hline
\hline
$\gamma$ & $\widetilde{m}=4$ & analytic & $\delta Re$ & $\delta Im$ &$\gamma$ & Prony fit & analytic &  $\delta Re$ & $\delta Im$ \\
\hline
$0$ & $0.292930-0.097663 i$ & $0.292833-0.097732 i$ & $0.033\%$ & $0.071\%$  &   $0$ & $0.292945-0.097662 i$ & $0.292833-0.097732 i$ & $0.038\%$ & $0.071\%$\\
$0.3$ & $0.296675-0.095014 i$ & $0.296425-0.094932 i$ & $0.084\%$ & $0.086\%$ &    $0.2$ & $0.295415-0.095958 i$ & $0.295203-0.095927 i$ & $0.072\%$ & $0.033\%$\\
$0.5$ & $0.299268-0.092873 i$ & $0.298948-0.092732 i$ & $0.107\%$ & $0.152\%$&       $0.4$ & $0.297973-0.093981 i$ & $0.297674-0.093869 i$ & $0.100\%$ & $0.119\%$\\
$0.7$ & $0.301894-0.090341 i$ & $0.301572-0.090212 i$ & $0.107\%$ & $0.143\%$&        $0.6$ & $0.300593-0.091654 i$ & $0.300247-0.091515 i$ & $0.115\%$ & $0.152\%$\\
$0.9$ & $0.304448-0.087316 i$ & $0.304300-0.087327 i$ & $0.049\%$ & $0.013\%$&     $0.8$ & $0.303205-0.088879 i$ & $0.302923-0.088818 i$ & $0.093\%$ & $0.069\%$\\
$1.1$ & $0.306703-0.083742 i$ & $0.307130-0.084032 i$ & $0.139\%$ & $0.347\%$&        $1.$ & $0.305640-0.085560 i$ & $0.305702-0.085734 i$ & $0.020\%$ & $0.203\%$\\
$1.18$ & $0.307461-0.082194 i$ & $0.308291-0.082590 i$ & $0.270\%$ & $0.482\%$&       $1.18$ & $0.307442-0.082165 i$ & $0.308291-0.082590 i$ & $0.276\%$ & $0.517\%$\\
\hline
\hline
\end{tabular}
}
\caption{QNMs for the $\ell=1$ ($n=0$) scalar field for the Hayward BH obtained by the analytic formula,  6th order WKB formula with Padé approximants  and time-domain integration together with relative differences $\delta Re$ and $\delta Im$ for real and imaginary parts for various $\gamma$ and $M=1$.}
\end{table*}

\begin{table*}
\centering
\resizebox{\textwidth}{!}{
\begin{tabular}{c c c c c c c c c c}
\hline
\hline
$\gamma$ & $\widetilde{m}=4$ & analytic &  $\delta Re$ & $\delta Im$ & $\gamma$ & Prony fit & analytic &  $\delta Re$ & $\delta Im$  \\
\hline
$0$ & $0.182643-0.096566 i$ & $0.182649-0.096943 i$ & $0.004\%$ & $0.390\%$&      $0$ & $0.183033-0.096909 i$ & $0.182649-0.096943 i$ & $0.209\%$ & $0.035\%$\\
$0.3$ & $0.185468-0.093598 i$ & $0.184860-0.093768 i$ & $0.328\%$ & $0.182\%$&      $0.2$ & $0.184667-0.094953 i$ & $0.184120-0.094899 i$ & $0.296\%$ & $0.056\%$\\
$0.5$ & $0.187134-0.091198 i$ & $0.186343-0.091253 i$ & $0.423\%$ & $0.050\%$&     $0.4$ & $0.186304-0.092685 i$ & $0.185602-0.092555 i$ & $0.377\%$ & $0.141\%$\\
$0.7$ & $0.188670-0.088379 i$ & $0.187812-0.088355 i$ & $0.455\%$ & $0.026\%$&     $0.6$ & $0.187875-0.090012 i$ & $0.187080-0.089856 i$ & $0.423\%$ & $0.174\%$\\
$0.9$ & $0.189803-0.084872 i$ & $0.189252-0.085020 i$ & $0.290\%$ & $0.175\%$&      $0.8$ & $0.189213-0.086799 i$ & $0.188537-0.086746 i$ & $0.358\%$ & $0.061\%$\\
$1.1$ & $0.189978-0.080924 i$ & $0.190645-0.081192 i$ & $0.351\%$ & $0.331\%$&      $1.$ & $0.189919-0.082984 i$ & $0.189955-0.083171 i$ & $0.0192\%$ & $0.225\%$\\
$1.18$ & $0.189773-0.079501 i$ & $0.191185-0.079511 i$ & $0.744\%$ & $0.0123\%$&    $1.18$ & $0.189608-0.079548 i$ & $0.191185-0.079511 i$ & $0.832\%$ & $0.048\%$\\
\hline
\hline
\end{tabular}
}
\caption{QNMs for the $\ell=1/2$ ($n=0$) Dirac field for the Hayward BH obtained by the analytic formula and the 6th order WKB formula with Padé approximants and time-domain integration together with relative differences $\delta Re$ and $\delta Im$ for real and imaginary parts  for various $\gamma$ and $M=1$.}
\end{table*}

\begin{table*}
\begin{tabular}{c c c c c}
\hline
\hline
$\gamma$ & $\widetilde{m}=4$ & analytic & r.e. $Re (\omega)$ & r.e. $Im (\omega)$  \\
\hline
$0$ & $0.248255-0.092497 i$ & $0.250066-0.092980 i$ & $0.730\%$ & $0.523\%$\\
$0.3$ & $0.252975-0.090080 i$ & $0.254190-0.090142 i$ & $0.480\%$ & $0.0694\%$\\
$0.5$ & $0.256361-0.088012 i$ & $0.257137-0.087862 i$ & $0.303\%$ & $0.170\%$\\
$0.7$ & $0.259918-0.085437 i$ & $0.260256-0.085206 i$ & $0.130\%$ & $0.271\%$\\
$0.9$ & $0.263546-0.082152 i$ & $0.263556-0.082116 i$ & $0.00378\%$ & $0.0438\%$\\
$1.1$ & $0.266932-0.077924 i$ & $0.267047-0.078538 i$ & $0.0432\%$ & $0.788\%$\\
$1.18$ & $0.268083-0.075994 i$ & $0.268499-0.076956 i$ & $0.155\%$ & $1.27\%$\\
\hline
\hline
\end{tabular}
\caption{QNMs for the $\ell=1$ ($n=0$) Maxwell field for the Hayward BH obtained by the analytic formula and the 6th order WKB formula with Padé  approximants for various  $\gamma$ and $M=1$.}
\end{table*}


\textbf{Eikonal formula and analytic expression beyond eikonal.} The wave-like equation can be  approximated by the following expression:
\begin{equation}\label{potential-multipole}
V(r_*)=\kappa^2\left(H(r_*)+\Order{\kappa^{-1}}\right),
\end{equation}
where $\kappa\equiv\ell+\frac{1}{2}$ and $\ell=s,s+1,s+2,\ldots$. 
In this work we will use the approach developed in \cite{Konoplya:2023moy} and applied by the author in \cite{Malik:2024voy,Malik:2024sxv} and expand the position of the potential's maximum and the frequency  in terms of $\kappa^{-1}$.
The function $H(r_*)$ possesses one maximum, and its position  (\ref{potential-multipole}) can be written as follows:
\begin{equation}\label{rmax}
  r_{\max }=r_0+r_1\kappa^{-1}+r_2\kappa^{-2}+\ldots.
\end{equation}

Using (\ref{rmax}) into the following first order WKB formula for the frequency
\begin{eqnarray}
\omega&=&\sqrt{V_0-\imo \K\sqrt{-2V_2}},
\end{eqnarray}
and further expanding it in terms of powers of $\kappa^{-1}$, one finds that \cite{Konoplya:2023moy}
\begin{eqnarray}\label{eikonal-formulas}
\omega=\Omega\kappa-\imo\lambda\K+\Order{\kappa^{-1}}.
\end{eqnarray}
The obtained expression serves as an approximation for $\kappa\gg\K$. This approach has been recently used by the author to find sufficiently accurate and compact analytic expressions for QNMs of dilaton \cite{Malik:2024sxv} and Reissner-Nordstrom-like BHs \cite{Malik:2024voy}.

Cardoso et. al. claimed \cite{Cardoso:2008bp}  that the unstable null geodesics around a static, spherically symmetric and asymptotically flat or de Sitter BH are linked to its quasinormal frequencies in the regime where $\ell \gg n$: the real and imaginary components of the quasinormal frequency are proportional to the rotational frequency $\Omega$ and Lyapunov exponent $\lambda$ of the circular null geodesics. This relationship is expressed by the following equation \cite{Cardoso:2008bp}:
\begin{equation}\label{QNM}
\omega_n=\Omega\ell-\imo(n+1/2)|\lambda|, \quad \ell \rightarrow \infty.
\end{equation}

The aforementioned correspondence takes place for the Schwarzschild BH and some its generalizations. Later, in \cite{Konoplya:2017wot} it was proved that it breaks down provided the standard centrifugal term, $f(r) \ell (\ell +1)/r^2$, in the wave equation takes on a different form, as it happens, for instance, in the Einstein-(dilaton)-Gauss-Bonnet and Einstein-Lovelock theories \cite{Konoplya:2017wot,Konoplya:2020bxa,Konoplya:2019hml}. Moreover, for asymptotically de Sitter BHs the overtones number $n$ is not anymore the actual frequency number \cite{Konoplya:2022gjp}, because the spectrum in that case consists from the two branches, the Schwarzschild branch and the de Sitter branch. The latter branch cannot be found by the WKB method, which is the reason for the breakdown of the correspondence.

From tables I - VI we can see that the as long as $\ell >0$ the obtained analytical formulas are quite precise and the relative error is much less than one percent for the whole range of the parameter $\gamma$. Such small relative errors are confirmed by the comparison with both 6th order WKB approach with Padé approximants and with the time-domain integration. For $\ell =0$ this is not so and the relative error could be considerable reaching several percents.

The precision of the obtained analytic formulas for QNMs can be checked by comparison with the two methods:  the 6th order WKB formula with Padé approximants and,  the time-domain integration. While the first way is not fully independent, because it is a modification of the WKB method, the second method  is fully independent from the WKB approach.
Here for the integration in time domain we used the Gundlach-Price-Pullin discretization scheme \cite{Gundlach:1993tp}
$\Psi\left(N\right)=\Psi\left(W\right)+\Psi\left(E\right)-\Psi\left(S\right)
- \Delta^2V\left(S\right)\frac{\Psi\left(W\right)+\Psi\left(E\right)}{8}+{\cal O}\left(\Delta^4\right).$
Here, we used the following designations for the points: $N\equiv\left(u+\Delta,v+\Delta\right)$, $W\equiv\left(u+\Delta,v\right)$, $E\equiv\left(u,v+\Delta\right)$, and $S\equiv\left(u,v\right)$. This method was used a in numerous publications \cite{Konoplya:2014lha,Ishihara:2008re,Churilova:2021tgn,Qian:2022kaq,Aneesh:2018hlp,Varghese:2011ku} and showed great concordance for the lowest lying modes with other methods, such as the Leaver method.

Here we were limited by the third order in powers of $\gamma$ and a first few orders beyond eikonal limit.  Similarly, the expansion (\ref{rmax}) can be done up to any desired order at least in principle. One can find also the higher-order corrections to the eikonal formula (\ref{eikonal-formulas}), where the $n-th$ order WKB formula can be used, in order to find the correction of the order $\kappa^{1-n}$ to the analytic approximation (\ref{eikonal-formulas}).

\begin{widetext}
Expanding in powers of $\kappa^{-1}$, similarly to \cite{Konoplya:2023moy}, we obtain the expansion for the position of the potential maximum for a massless scalar field,
\begin{equation}\label{rmax-scalar}
\begin{array}{rcl}
r_{\max } &=& \displaystyle3 M-\frac{M}{3 \kappa ^2}+\gamma  \left(\frac{4 M}{27 \kappa ^2}-\frac{2 M}{9}\right)+\gamma ^2 \left(\frac{23 M}{729 \kappa ^2}-\frac{M}{27}\right)+\gamma ^3 \left(\frac{226 M}{19683 \kappa ^2}-\frac{70 M}{6561}\right)+\mathcal{O}\left(\gamma ^4,\frac{1}{\kappa ^4}\right).
\end{array}
\end{equation}
Then,  via the WKB formula, we obtain the expression for the frequency
\begin{equation}\label{eikonal-scalar}
\begin{array}{rcl}
\omega  &=& \displaystyle-\frac{i K \left(940 K^2+313\right)}{46656 \sqrt{3} M \kappa ^2}+\frac{29-60 K^2}{1296 \sqrt{3} M \kappa }+\frac{\kappa }{3 \sqrt{3} M}-\frac{i K}{3 \sqrt{3} M}\\
&&\displaystyle+\gamma  \left(\frac{5 i K \left(164 K^2+179\right)}{104976 \sqrt{3} M \kappa ^2}+\frac{420 K^2-23}{34992 \sqrt{3} M \kappa }+\frac{\kappa }{81 \sqrt{3} M}+\frac{2 i K}{81 \sqrt{3} M}\right)\\
&&\displaystyle+\gamma ^2 \left(\frac{i K \left(2300 K^2+587\right)}{708588 \sqrt{3} M \kappa ^2}-\frac{3756 K^2+1583}{1889568 \sqrt{3} M \kappa }+\frac{\kappa }{486 \sqrt{3} M}+\frac{2 i K}{243 \sqrt{3} M}\right)\\
&&\displaystyle+\gamma ^3 \left(\frac{i K \left(331220 K^2+47879\right)}{306110016 \sqrt{3} M \kappa ^2}-\frac{124764 K^2+27199}{51018336 \sqrt{3} M \kappa }+\frac{61 \kappa }{118098 \sqrt{3} M}+\frac{181 i K}{59049 \sqrt{3} M}\right)+\mathcal{O}\left(\gamma ^4,\frac{1}{\kappa ^3}\right).
\end{array}
\end{equation}
Here we have  $\kappa\equiv\ell+1/2$, $K\equiv n+1/2$.
The potential peak expansion for the electromagnetic field has the form
\begin{equation}\label{rmax-electromagnetic}
r_{\max } = 3 M-\frac{2 \gamma  M}{9}-\frac{\gamma ^2 M}{27}-\frac{70 \gamma ^3 M}{6561}+\mathcal{O}\left(\gamma ^4,\frac{1}{\kappa ^4}\right)
\end{equation}
and the analytic expression for the QNMs has the form
\begin{equation}\label{eikonal-electromagnetic}
\begin{array}{rcl}
\omega  &=& \displaystyle-\frac{5 i K \left(188 K^2-283\right)}{46656 \sqrt{3} M \kappa ^2}-\frac{5 \left(12 K^2+23\right)}{1296 \sqrt{3} M \kappa }+\frac{\kappa }{3 \sqrt{3} M}-\frac{i K}{3 \sqrt{3} M}\\
&&\displaystyle+\gamma  \left(\frac{5 i K \left(164 K^2+179\right)}{104976 \sqrt{3} M \kappa ^2}+\frac{420 K^2+121}{34992 \sqrt{3} M \kappa }+\frac{\kappa }{81 \sqrt{3} M}+\frac{2 i K}{81 \sqrt{3} M}\right)\\
&&\displaystyle+\gamma ^2 \left(\frac{i K \left(2300 K^2+2531\right)}{708588 \sqrt{3} M \kappa ^2}+\frac{1153-3756 K^2}{1889568 \sqrt{3} M \kappa }+\frac{\kappa }{486 \sqrt{3} M}+\frac{2 i K}{243 \sqrt{3} M}\right)\\
&&\displaystyle+\gamma ^3 \left(\frac{i K \left(331220 K^2+616391\right)}{306110016 \sqrt{3} M \kappa ^2}-\frac{124764 K^2+559}{51018336 \sqrt{3} M \kappa }+\frac{61 \kappa }{118098 \sqrt{3} M}+\frac{181 i K}{59049 \sqrt{3} M}\right)+\mathcal{O}\left(\gamma ^4,\frac{1}{\kappa ^3}\right)
\end{array}
\end{equation}
For the Dirac field, the position of the maximum of the potential is
\begin{equation}\label{rmax-Dirac}
\begin{array}{rcl}
r_{\max } &=& \displaystyle\frac{11 M}{16 \sqrt{3} \kappa ^3}-\frac{\sqrt{3} M}{2 \kappa }+3 M+\gamma  \left(-\frac{59 M}{432 \sqrt{3} \kappa ^3}+\frac{M}{9 \kappa ^2}+\frac{M}{6 \sqrt{3} \kappa }-\frac{2 M}{9}\right)\\
&&\displaystyle+\gamma ^2 \left(-\frac{89 M}{2592 \sqrt{3} \kappa ^3}+\frac{19 M}{972 \kappa ^2}+\frac{M}{36 \sqrt{3} \kappa }-\frac{M}{27}\right)\\
&&\displaystyle+\gamma ^3 \left(-\frac{6047 M}{629856 \sqrt{3} \kappa ^3}+\frac{16 M}{2187 \kappa ^2}+\frac{211 M}{26244 \sqrt{3} \kappa }-\frac{70 M}{6561}\right)+\mathcal{O}\left(\gamma ^4,\frac{1}{\kappa ^4}\right)
\end{array}
\end{equation}
and the QNMs are
\begin{equation}\label{eikonal-Dirac}
\begin{array}{rcl}
\omega  &=& \displaystyle\frac{i K \left(119-940 K^2\right)}{46656 \sqrt{3} M \kappa ^2}-\frac{60 K^2+7}{1296 \sqrt{3} M \kappa }+\frac{\kappa }{3 \sqrt{3} M}-\frac{i K}{3 \sqrt{3} M}\\
&&\displaystyle+\gamma  \left(\frac{i K \left(820 K^2+679\right)}{104976 \sqrt{3} M \kappa ^2}+\frac{5 \left(84 K^2-19\right)}{34992 \sqrt{3} M \kappa }+\frac{\kappa }{81 \sqrt{3} M}+\frac{2 i K}{81 \sqrt{3} M}\right)\\
&&\displaystyle+\gamma ^2 \left(\frac{23 i K \left(100 K^2+49\right)}{708588 \sqrt{3} M \kappa ^2}-\frac{3756 K^2+1763}{1889568 \sqrt{3} M \kappa }+\frac{\kappa }{486 \sqrt{3} M}+\frac{2 i K}{243 \sqrt{3} M}\right)\\
&&\displaystyle+\gamma ^3 \left(\frac{5 i K \left(66244 K^2+38347\right)}{306110016 \sqrt{3} M \kappa ^2}-\frac{124764 K^2+26047}{51018336 \sqrt{3} M \kappa }+\frac{61 \kappa }{118098 \sqrt{3} M}+\frac{181 i K}{59049 \sqrt{3} M}\right)+\mathcal{O}\left(\gamma ^4,\frac{1}{\kappa ^3}\right)
\end{array}
\end{equation}
\end{widetext}

From the above analytical expressions and more accurate numerical data, we observe that the real oscillation frequency, represented by $Re(\omega)$, increases with the coupling constant $\gamma$, while the damping rate, given by $Im(\omega)$, decreases as $\gamma$ becomes larger. Consequently, the quality factor, which is proportional to the ratio of these two quantities, increases when $\gamma$ is introduced. This implies that a quantum-corrected black hole is a better oscillator than its classical counterpart.

\textbf{Conclusions.} The Hayward metric is important as it provides a regular BH solution that avoids singularities, offering a more complete and realistic model of BHs within the framework of general relativity. This metric also facilitates the study of quantum effects near BHs, bridging the gap between classical and quantum gravity theories.

In the present research we have derived compact analytical approximate expressions  for QNMs of the Hayward BH \cite{Berglund:2012bu} for scalar, Dirac and Maxwell perturbations. Comparison of the results obtained by the analytic formula with those found by the 6th order WKB formula using Padé approximants  and time-domain integration methods demonstrate remarkable precision of the analytic formula: the relative error is much smaller than one percent for the whole range of $\gamma$ and $\ell >0$.

Using the same approach one could obtain analytic expressions for the grey-body factors, which could be interesting in view of the recent correspondences between the profile of high frequency waves, QNMs and grey-body factors  \cite{Oshita:2023cjz,Konoplya:2024lir,Rosato:2024arw}.

\textbf{Acknowledgments.} The author thanks R. A. Konoplya for stimulating discussions and careful reading of the manuscript.


\bibliographystyle{apsrev4-1}
\bibliography{bibliography}

\begin{thebibliography}{81}%
\makeatletter
\providecommand \@ifxundefined [1]{%
 \@ifx{#1\undefined}
}%
\providecommand \@ifnum [1]{%
 \ifnum #1\expandafter \@firstoftwo
 \else \expandafter \@secondoftwo
 \fi
}%
\providecommand \@ifx [1]{%
 \ifx #1\expandafter \@firstoftwo
 \else \expandafter \@secondoftwo
 \fi
}%
\providecommand \natexlab [1]{#1}%
\providecommand \enquote  [1]{``#1''}%
\providecommand \bibnamefont  [1]{#1}%
\providecommand \bibfnamefont [1]{#1}%
\providecommand \citenamefont [1]{#1}%
\providecommand \href@noop [0]{\@secondoftwo}%
\providecommand \href [0]{\begingroup \@sanitize@url \@href}%
\providecommand \@href[1]{\@@startlink{#1}\@@href}%
\providecommand \@@href[1]{\endgroup#1\@@endlink}%
\providecommand \@sanitize@url [0]{\catcode `\\12\catcode `\$12\catcode
  `\&12\catcode `\#12\catcode `\^12\catcode `\_12\catcode `\%12\relax}%
\providecommand \@@startlink[1]{}%
\providecommand \@@endlink[0]{}%
\providecommand \url  [0]{\begingroup\@sanitize@url \@url }%
\providecommand \@url [1]{\endgroup\@href {#1}{\urlprefix }}%
\providecommand \urlprefix  [0]{URL }%
\providecommand \Eprint [0]{\href }%
\providecommand \doibase [0]{http://dx.doi.org/}%
\providecommand \selectlanguage [0]{\@gobble}%
\providecommand \bibinfo  [0]{\@secondoftwo}%
\providecommand \bibfield  [0]{\@secondoftwo}%
\providecommand \translation [1]{[#1]}%
\providecommand \BibitemOpen [0]{}%
\providecommand \bibitemStop [0]{}%
\providecommand \bibitemNoStop [0]{.\EOS\space}%
\providecommand \EOS [0]{\spacefactor3000\relax}%
\providecommand \BibitemShut  [1]{\csname bibitem#1\endcsname}%
\let\auto@bib@innerbib\@empty
\bibitem [{\citenamefont {Abbott}\ \emph {et~al.}(2016)\citenamefont {Abbott}
  \emph {et~al.}}]{LIGOScientific:2016aoc}%
  \BibitemOpen
  \bibfield  {author} {\bibinfo {author} {\bibfnamefont {B.~P.}\ \bibnamefont
  {Abbott}} \emph {et~al.} (\bibinfo {collaboration} {LIGO Scientific,
  Virgo}),\ }\href {\doibase 10.1103/PhysRevLett.116.061102} {\bibfield
  {journal} {\bibinfo  {journal} {Phys. Rev. Lett.}\ }\textbf {\bibinfo
  {volume} {116}},\ \bibinfo {pages} {061102} (\bibinfo {year} {2016})},\
  \Eprint {http://arxiv.org/abs/1602.03837} {arXiv:1602.03837 [gr-qc]}
  \BibitemShut {NoStop}%
\bibitem [{\citenamefont {Akiyama}\ \emph {et~al.}(2019)\citenamefont {Akiyama}
  \emph {et~al.}}]{EventHorizonTelescope:2019dse}%
  \BibitemOpen
  \bibfield  {author} {\bibinfo {author} {\bibfnamefont {K.}~\bibnamefont
  {Akiyama}} \emph {et~al.} (\bibinfo {collaboration} {Event Horizon
  Telescope}),\ }\href {\doibase 10.3847/2041-8213/ab0ec7} {\bibfield
  {journal} {\bibinfo  {journal} {Astrophys. J. Lett.}\ }\textbf {\bibinfo
  {volume} {875}},\ \bibinfo {pages} {L1} (\bibinfo {year} {2019})},\ \Eprint
  {http://arxiv.org/abs/1906.11238} {arXiv:1906.11238 [astro-ph.GA]}
  \BibitemShut {NoStop}%
\bibitem [{\citenamefont {Goddi}\ \emph {et~al.}(2016)\citenamefont {Goddi}
  \emph {et~al.}}]{Goddi:2016qax}%
  \BibitemOpen
  \bibfield  {author} {\bibinfo {author} {\bibfnamefont {C.}~\bibnamefont
  {Goddi}} \emph {et~al.},\ }\href {\doibase 10.1142/9789813226609_0046}
  {\bibfield  {journal} {\bibinfo  {journal} {Int. J. Mod. Phys. D}\ }\textbf
  {\bibinfo {volume} {26}},\ \bibinfo {pages} {1730001} (\bibinfo {year}
  {2016})},\ \Eprint {http://arxiv.org/abs/1606.08879} {arXiv:1606.08879
  [astro-ph.HE]} \BibitemShut {NoStop}%
\bibitem [{\citenamefont {Babak}\ \emph {et~al.}(2017)\citenamefont {Babak},
  \citenamefont {Gair}, \citenamefont {Sesana}, \citenamefont {Barausse},
  \citenamefont {Sopuerta}, \citenamefont {Berry}, \citenamefont {Berti},
  \citenamefont {Amaro-Seoane}, \citenamefont {Petiteau},\ and\ \citenamefont
  {Klein}}]{Babak:2017tow}%
  \BibitemOpen
  \bibfield  {author} {\bibinfo {author} {\bibfnamefont {S.}~\bibnamefont
  {Babak}}, \bibinfo {author} {\bibfnamefont {J.}~\bibnamefont {Gair}},
  \bibinfo {author} {\bibfnamefont {A.}~\bibnamefont {Sesana}}, \bibinfo
  {author} {\bibfnamefont {E.}~\bibnamefont {Barausse}}, \bibinfo {author}
  {\bibfnamefont {C.~F.}\ \bibnamefont {Sopuerta}}, \bibinfo {author}
  {\bibfnamefont {C.~P.~L.}\ \bibnamefont {Berry}}, \bibinfo {author}
  {\bibfnamefont {E.}~\bibnamefont {Berti}}, \bibinfo {author} {\bibfnamefont
  {P.}~\bibnamefont {Amaro-Seoane}}, \bibinfo {author} {\bibfnamefont
  {A.}~\bibnamefont {Petiteau}}, \ and\ \bibinfo {author} {\bibfnamefont
  {A.}~\bibnamefont {Klein}},\ }\href {\doibase 10.1103/PhysRevD.95.103012}
  {\bibfield  {journal} {\bibinfo  {journal} {Phys. Rev. D}\ }\textbf {\bibinfo
  {volume} {95}},\ \bibinfo {pages} {103012} (\bibinfo {year} {2017})},\
  \Eprint {http://arxiv.org/abs/1703.09722} {arXiv:1703.09722 [gr-qc]}
  \BibitemShut {NoStop}%
\bibitem [{\citenamefont {Tsukamoto}\ \emph {et~al.}(2014)\citenamefont
  {Tsukamoto}, \citenamefont {Li},\ and\ \citenamefont
  {Bambi}}]{Tsukamoto:2014tja}%
  \BibitemOpen
  \bibfield  {author} {\bibinfo {author} {\bibfnamefont {N.}~\bibnamefont
  {Tsukamoto}}, \bibinfo {author} {\bibfnamefont {Z.}~\bibnamefont {Li}}, \
  and\ \bibinfo {author} {\bibfnamefont {C.}~\bibnamefont {Bambi}},\ }\href
  {\doibase 10.1088/1475-7516/2014/06/043} {\bibfield  {journal} {\bibinfo
  {journal} {JCAP}\ }\textbf {\bibinfo {volume} {06}},\ \bibinfo {pages} {043}
  (\bibinfo {year} {2014})},\ \Eprint {http://arxiv.org/abs/1403.0371}
  {arXiv:1403.0371 [gr-qc]} \BibitemShut {NoStop}%
\bibitem [{\citenamefont {Shaikh}\ \emph {et~al.}(2021)\citenamefont {Shaikh},
  \citenamefont {Pal}, \citenamefont {Pal},\ and\ \citenamefont
  {Sarkar}}]{Shaikh:2021yux}%
  \BibitemOpen
  \bibfield  {author} {\bibinfo {author} {\bibfnamefont {R.}~\bibnamefont
  {Shaikh}}, \bibinfo {author} {\bibfnamefont {K.}~\bibnamefont {Pal}},
  \bibinfo {author} {\bibfnamefont {K.}~\bibnamefont {Pal}}, \ and\ \bibinfo
  {author} {\bibfnamefont {T.}~\bibnamefont {Sarkar}},\ }\href {\doibase
  10.1093/mnras/stab1779} {\bibfield  {journal} {\bibinfo  {journal} {Mon. Not.
  Roy. Astron. Soc.}\ }\textbf {\bibinfo {volume} {506}},\ \bibinfo {pages}
  {1229} (\bibinfo {year} {2021})},\ \Eprint {http://arxiv.org/abs/2102.04299}
  {arXiv:2102.04299 [gr-qc]} \BibitemShut {NoStop}%
\bibitem [{\citenamefont {Bronnikov}\ and\ \citenamefont
  {Konoplya}(2020)}]{Bronnikov:2019sbx}%
  \BibitemOpen
  \bibfield  {author} {\bibinfo {author} {\bibfnamefont {K.~A.}\ \bibnamefont
  {Bronnikov}}\ and\ \bibinfo {author} {\bibfnamefont {R.~A.}\ \bibnamefont
  {Konoplya}},\ }\href {\doibase 10.1103/PhysRevD.101.064004} {\bibfield
  {journal} {\bibinfo  {journal} {Phys. Rev. D}\ }\textbf {\bibinfo {volume}
  {101}},\ \bibinfo {pages} {064004} (\bibinfo {year} {2020})},\ \Eprint
  {http://arxiv.org/abs/1912.05315} {arXiv:1912.05315 [gr-qc]} \BibitemShut
  {NoStop}%
\bibitem [{\citenamefont {Panotopoulos}\ and\ \citenamefont
  {Rincón}(2019)}]{Panotopoulos:2019qjk}%
  \BibitemOpen
  \bibfield  {author} {\bibinfo {author} {\bibfnamefont {G.}~\bibnamefont
  {Panotopoulos}}\ and\ \bibinfo {author} {\bibfnamefont {n.}~\bibnamefont
  {Rincón}},\ }\href {\doibase 10.1140/epjp/i2019-12686-x} {\bibfield
  {journal} {\bibinfo  {journal} {Eur. Phys. J. Plus}\ }\textbf {\bibinfo
  {volume} {134}},\ \bibinfo {pages} {300} (\bibinfo {year} {2019})},\ \Eprint
  {http://arxiv.org/abs/1904.10847} {arXiv:1904.10847 [gr-qc]} \BibitemShut
  {NoStop}%
\bibitem [{\citenamefont {Yang}\ \emph {et~al.}(2021)\citenamefont {Yang},
  \citenamefont {Liu}, \citenamefont {Xu}, \citenamefont {Xing}, \citenamefont
  {Wu},\ and\ \citenamefont {Long}}]{Yang:2021cvh}%
  \BibitemOpen
  \bibfield  {author} {\bibinfo {author} {\bibfnamefont {Y.}~\bibnamefont
  {Yang}}, \bibinfo {author} {\bibfnamefont {D.}~\bibnamefont {Liu}}, \bibinfo
  {author} {\bibfnamefont {Z.}~\bibnamefont {Xu}}, \bibinfo {author}
  {\bibfnamefont {Y.}~\bibnamefont {Xing}}, \bibinfo {author} {\bibfnamefont
  {S.}~\bibnamefont {Wu}}, \ and\ \bibinfo {author} {\bibfnamefont {Z.-W.}\
  \bibnamefont {Long}},\ }\href {\doibase 10.1103/PhysRevD.104.104021}
  {\bibfield  {journal} {\bibinfo  {journal} {Phys. Rev. D}\ }\textbf {\bibinfo
  {volume} {104}},\ \bibinfo {pages} {104021} (\bibinfo {year} {2021})},\
  \Eprint {http://arxiv.org/abs/2107.06554} {arXiv:2107.06554 [gr-qc]}
  \BibitemShut {NoStop}%
\bibitem [{\citenamefont {Wahlang}\ \emph {et~al.}(2017)\citenamefont
  {Wahlang}, \citenamefont {Jeena},\ and\ \citenamefont
  {Chakrabarti}}]{Wahlang:2017zvk}%
  \BibitemOpen
  \bibfield  {author} {\bibinfo {author} {\bibfnamefont {W.}~\bibnamefont
  {Wahlang}}, \bibinfo {author} {\bibfnamefont {P.~A.}\ \bibnamefont {Jeena}},
  \ and\ \bibinfo {author} {\bibfnamefont {S.}~\bibnamefont {Chakrabarti}},\
  }\href {\doibase 10.1142/S0218271817501607} {\bibfield  {journal} {\bibinfo
  {journal} {Int. J. Mod. Phys. D}\ }\textbf {\bibinfo {volume} {26}},\
  \bibinfo {pages} {1750160} (\bibinfo {year} {2017})},\ \Eprint
  {http://arxiv.org/abs/1703.04286} {arXiv:1703.04286 [gr-qc]} \BibitemShut
  {NoStop}%
\bibitem [{\citenamefont {Saleh}\ \emph {et~al.}(2018)\citenamefont {Saleh},
  \citenamefont {Thomas},\ and\ \citenamefont {Kofane}}]{Saleh:2018hba}%
  \BibitemOpen
  \bibfield  {author} {\bibinfo {author} {\bibfnamefont {M.}~\bibnamefont
  {Saleh}}, \bibinfo {author} {\bibfnamefont {B.~B.}\ \bibnamefont {Thomas}}, \
  and\ \bibinfo {author} {\bibfnamefont {T.~C.}\ \bibnamefont {Kofane}},\
  }\href {\doibase 10.1140/epjc/s10052-018-5818-9} {\bibfield  {journal}
  {\bibinfo  {journal} {Eur. Phys. J. C}\ }\textbf {\bibinfo {volume} {78}},\
  \bibinfo {pages} {325} (\bibinfo {year} {2018})}\BibitemShut {NoStop}%
\bibitem [{\citenamefont {Toshmatov}\ \emph {et~al.}(2015)\citenamefont
  {Toshmatov}, \citenamefont {Abdujabbarov}, \citenamefont {Stuchlík},\ and\
  \citenamefont {Ahmedov}}]{Toshmatov:2015wga}%
  \BibitemOpen
  \bibfield  {author} {\bibinfo {author} {\bibfnamefont {B.}~\bibnamefont
  {Toshmatov}}, \bibinfo {author} {\bibfnamefont {A.}~\bibnamefont
  {Abdujabbarov}}, \bibinfo {author} {\bibfnamefont {Z.}~\bibnamefont
  {Stuchlík}}, \ and\ \bibinfo {author} {\bibfnamefont {B.}~\bibnamefont
  {Ahmedov}},\ }\href {\doibase 10.1103/PhysRevD.91.083008} {\bibfield
  {journal} {\bibinfo  {journal} {Phys. Rev. D}\ }\textbf {\bibinfo {volume}
  {91}},\ \bibinfo {pages} {083008} (\bibinfo {year} {2015})},\ \Eprint
  {http://arxiv.org/abs/1503.05737} {arXiv:1503.05737 [gr-qc]} \BibitemShut
  {NoStop}%
\bibitem [{\citenamefont {Wu}\ \emph {et~al.}(2022)\citenamefont {Wu},
  \citenamefont {Wang}, \citenamefont {Liu},\ and\ \citenamefont
  {Long}}]{Wu:2022eiv}%
  \BibitemOpen
  \bibfield  {author} {\bibinfo {author} {\bibfnamefont {S.~R.}\ \bibnamefont
  {Wu}}, \bibinfo {author} {\bibfnamefont {B.~Q.}\ \bibnamefont {Wang}},
  \bibinfo {author} {\bibfnamefont {D.}~\bibnamefont {Liu}}, \ and\ \bibinfo
  {author} {\bibfnamefont {Z.~W.}\ \bibnamefont {Long}},\ }\href {\doibase
  10.1140/epjc/s10052-022-10938-1} {\bibfield  {journal} {\bibinfo  {journal}
  {Eur. Phys. J. C}\ }\textbf {\bibinfo {volume} {82}},\ \bibinfo {pages} {998}
  (\bibinfo {year} {2022})},\ \Eprint {http://arxiv.org/abs/2201.08415}
  {arXiv:2201.08415 [gr-qc]} \BibitemShut {NoStop}%
\bibitem [{\citenamefont {Xi}\ and\ \citenamefont {Ao}(2016)}]{Xi:2016qrg}%
  \BibitemOpen
  \bibfield  {author} {\bibinfo {author} {\bibfnamefont {P.}~\bibnamefont
  {Xi}}\ and\ \bibinfo {author} {\bibfnamefont {X.-c.}\ \bibnamefont {Ao}},\
  }\href {\doibase 10.1007/s10714-016-2017-6} {\bibfield  {journal} {\bibinfo
  {journal} {Gen. Rel. Grav.}\ }\textbf {\bibinfo {volume} {48}},\ \bibinfo
  {pages} {14} (\bibinfo {year} {2016})}\BibitemShut {NoStop}%
\bibitem [{\citenamefont {Jusufi}\ \emph {et~al.}(2021)\citenamefont {Jusufi},
  \citenamefont {Azreg-Aïnou}, \citenamefont {Jamil}, \citenamefont {Wei},
  \citenamefont {Wu},\ and\ \citenamefont {Wang}}]{Jusufi:2020odz}%
  \BibitemOpen
  \bibfield  {author} {\bibinfo {author} {\bibfnamefont {K.}~\bibnamefont
  {Jusufi}}, \bibinfo {author} {\bibfnamefont {M.}~\bibnamefont
  {Azreg-Aïnou}}, \bibinfo {author} {\bibfnamefont {M.}~\bibnamefont {Jamil}},
  \bibinfo {author} {\bibfnamefont {S.-W.}\ \bibnamefont {Wei}}, \bibinfo
  {author} {\bibfnamefont {Q.}~\bibnamefont {Wu}}, \ and\ \bibinfo {author}
  {\bibfnamefont {A.}~\bibnamefont {Wang}},\ }\href {\doibase
  10.1103/PhysRevD.103.024013} {\bibfield  {journal} {\bibinfo  {journal}
  {Phys. Rev. D}\ }\textbf {\bibinfo {volume} {103}},\ \bibinfo {pages}
  {024013} (\bibinfo {year} {2021})},\ \Eprint
  {http://arxiv.org/abs/2008.08450} {arXiv:2008.08450 [gr-qc]} \BibitemShut
  {NoStop}%
\bibitem [{\citenamefont {Guo}\ \emph {et~al.}(2024)\citenamefont {Guo},
  \citenamefont {Xie},\ and\ \citenamefont {Miao}}]{Guo:2024jhg}%
  \BibitemOpen
  \bibfield  {author} {\bibinfo {author} {\bibfnamefont {Y.}~\bibnamefont
  {Guo}}, \bibinfo {author} {\bibfnamefont {H.}~\bibnamefont {Xie}}, \ and\
  \bibinfo {author} {\bibfnamefont {Y.-G.}\ \bibnamefont {Miao}},\ }\href@noop
  {} {\  (\bibinfo {year} {2024})},\ \Eprint {http://arxiv.org/abs/2402.10406}
  {arXiv:2402.10406 [gr-qc]} \BibitemShut {NoStop}%
\bibitem [{\citenamefont {Franzin}\ \emph {et~al.}(2022)\citenamefont
  {Franzin}, \citenamefont {Liberati}, \citenamefont {Mazza}, \citenamefont
  {Dey},\ and\ \citenamefont {Chakraborty}}]{Franzin:2022iai}%
  \BibitemOpen
  \bibfield  {author} {\bibinfo {author} {\bibfnamefont {E.}~\bibnamefont
  {Franzin}}, \bibinfo {author} {\bibfnamefont {S.}~\bibnamefont {Liberati}},
  \bibinfo {author} {\bibfnamefont {J.}~\bibnamefont {Mazza}}, \bibinfo
  {author} {\bibfnamefont {R.}~\bibnamefont {Dey}}, \ and\ \bibinfo {author}
  {\bibfnamefont {S.}~\bibnamefont {Chakraborty}},\ }\href {\doibase
  10.1103/PhysRevD.105.124051} {\bibfield  {journal} {\bibinfo  {journal}
  {Phys. Rev. D}\ }\textbf {\bibinfo {volume} {105}},\ \bibinfo {pages}
  {124051} (\bibinfo {year} {2022})},\ \Eprint
  {http://arxiv.org/abs/2201.01650} {arXiv:2201.01650 [gr-qc]} \BibitemShut
  {NoStop}%
\bibitem [{\citenamefont {Al-Badawi}\ and\ \citenamefont
  {Jawad}(2024)}]{Al-Badawi:2024mco}%
  \BibitemOpen
  \bibfield  {author} {\bibinfo {author} {\bibfnamefont {A.}~\bibnamefont
  {Al-Badawi}}\ and\ \bibinfo {author} {\bibfnamefont {A.}~\bibnamefont
  {Jawad}},\ }\href {\doibase 10.1140/epjc/s10052-024-12478-2} {\bibfield
  {journal} {\bibinfo  {journal} {Eur. Phys. J. C}\ }\textbf {\bibinfo {volume}
  {84}},\ \bibinfo {pages} {115} (\bibinfo {year} {2024})}\BibitemShut
  {NoStop}%
\bibitem [{\citenamefont {Hayward}(2006)}]{Hayward:2005gi}%
  \BibitemOpen
  \bibfield  {author} {\bibinfo {author} {\bibfnamefont {S.~A.}\ \bibnamefont
  {Hayward}},\ }\href {\doibase 10.1103/PhysRevLett.96.031103} {\bibfield
  {journal} {\bibinfo  {journal} {Phys. Rev. Lett.}\ }\textbf {\bibinfo
  {volume} {96}},\ \bibinfo {pages} {031103} (\bibinfo {year} {2006})},\
  \Eprint {http://arxiv.org/abs/gr-qc/0506126} {arXiv:gr-qc/0506126}
  \BibitemShut {NoStop}%
\bibitem [{\citenamefont {Dutta~Roy}\ and\ \citenamefont
  {Kar}(2022)}]{DuttaRoy:2022ytr}%
  \BibitemOpen
  \bibfield  {author} {\bibinfo {author} {\bibfnamefont {P.}~\bibnamefont
  {Dutta~Roy}}\ and\ \bibinfo {author} {\bibfnamefont {S.}~\bibnamefont
  {Kar}},\ }\href {\doibase 10.1103/PhysRevD.106.044028} {\bibfield  {journal}
  {\bibinfo  {journal} {Phys. Rev. D}\ }\textbf {\bibinfo {volume} {106}},\
  \bibinfo {pages} {044028} (\bibinfo {year} {2022})},\ \Eprint
  {http://arxiv.org/abs/2206.04505} {arXiv:2206.04505 [gr-qc]} \BibitemShut
  {NoStop}%
\bibitem [{\citenamefont {Al-Badawi}\ and\ \citenamefont
  {Kraishan}(2024)}]{Al-Badawi:2023lke}%
  \BibitemOpen
  \bibfield  {author} {\bibinfo {author} {\bibfnamefont {A.}~\bibnamefont
  {Al-Badawi}}\ and\ \bibinfo {author} {\bibfnamefont {A.}~\bibnamefont
  {Kraishan}},\ }\href {\doibase 10.1016/j.cjph.2023.10.048} {\bibfield
  {journal} {\bibinfo  {journal} {Chin. J. Phys.}\ }\textbf {\bibinfo {volume}
  {87}},\ \bibinfo {pages} {59} (\bibinfo {year} {2024})}\BibitemShut {NoStop}%
\bibitem [{\citenamefont {Pedraza}\ \emph {et~al.}(2022)\citenamefont
  {Pedraza}, \citenamefont {López}, \citenamefont {Arceo},\ and\ \citenamefont
  {Cabrera-Munguia}}]{Pedraza:2021hzw}%
  \BibitemOpen
  \bibfield  {author} {\bibinfo {author} {\bibfnamefont {O.}~\bibnamefont
  {Pedraza}}, \bibinfo {author} {\bibfnamefont {L.~A.}\ \bibnamefont {López}},
  \bibinfo {author} {\bibfnamefont {R.}~\bibnamefont {Arceo}}, \ and\ \bibinfo
  {author} {\bibfnamefont {I.}~\bibnamefont {Cabrera-Munguia}},\ }\href
  {\doibase 10.1142/S0217732322500572} {\bibfield  {journal} {\bibinfo
  {journal} {Mod. Phys. Lett. A}\ }\textbf {\bibinfo {volume} {37}},\ \bibinfo
  {pages} {2250057} (\bibinfo {year} {2022})},\ \Eprint
  {http://arxiv.org/abs/2111.06488} {arXiv:2111.06488 [gr-qc]} \BibitemShut
  {NoStop}%
\bibitem [{\citenamefont {Lin}\ \emph {et~al.}(2013)\citenamefont {Lin},
  \citenamefont {Li},\ and\ \citenamefont {Yang}}]{Lin:2013ofa}%
  \BibitemOpen
  \bibfield  {author} {\bibinfo {author} {\bibfnamefont {K.}~\bibnamefont
  {Lin}}, \bibinfo {author} {\bibfnamefont {J.}~\bibnamefont {Li}}, \ and\
  \bibinfo {author} {\bibfnamefont {S.}~\bibnamefont {Yang}},\ }\href {\doibase
  10.1007/s10773-013-1682-4} {\bibfield  {journal} {\bibinfo  {journal} {Int.
  J. Theor. Phys.}\ }\textbf {\bibinfo {volume} {52}},\ \bibinfo {pages} {3771}
  (\bibinfo {year} {2013})}\BibitemShut {NoStop}%
\bibitem [{\citenamefont {Mukohyama}\ \emph {et~al.}(2023)\citenamefont
  {Mukohyama}, \citenamefont {Takahashi}, \citenamefont {Tomikawa},\ and\
  \citenamefont {Yingcharoenrat}}]{Mukohyama:2023xyf}%
  \BibitemOpen
  \bibfield  {author} {\bibinfo {author} {\bibfnamefont {S.}~\bibnamefont
  {Mukohyama}}, \bibinfo {author} {\bibfnamefont {K.}~\bibnamefont
  {Takahashi}}, \bibinfo {author} {\bibfnamefont {K.}~\bibnamefont {Tomikawa}},
  \ and\ \bibinfo {author} {\bibfnamefont {V.}~\bibnamefont {Yingcharoenrat}},\
  }\href {\doibase 10.1088/1475-7516/2023/07/050} {\bibfield  {journal}
  {\bibinfo  {journal} {JCAP}\ }\textbf {\bibinfo {volume} {07}},\ \bibinfo
  {pages} {050} (\bibinfo {year} {2023})},\ \Eprint
  {http://arxiv.org/abs/2304.14304} {arXiv:2304.14304 [gr-qc]} \BibitemShut
  {NoStop}%
\bibitem [{\citenamefont {Konoplya}(2023{\natexlab{a}})}]{Konoplya:2023ppx}%
  \BibitemOpen
  \bibfield  {author} {\bibinfo {author} {\bibfnamefont {R.~A.}\ \bibnamefont
  {Konoplya}},\ }\href {\doibase 10.1088/1475-7516/2023/07/001} {\bibfield
  {journal} {\bibinfo  {journal} {JCAP}\ }\textbf {\bibinfo {volume} {07}},\
  \bibinfo {pages} {001} (\bibinfo {year} {2023}{\natexlab{a}})},\ \Eprint
  {http://arxiv.org/abs/2305.09187} {arXiv:2305.09187 [gr-qc]} \BibitemShut
  {NoStop}%
\bibitem [{\citenamefont {Held}\ \emph {et~al.}(2019)\citenamefont {Held},
  \citenamefont {Gold},\ and\ \citenamefont {Eichhorn}}]{Held:2019xde}%
  \BibitemOpen
  \bibfield  {author} {\bibinfo {author} {\bibfnamefont {A.}~\bibnamefont
  {Held}}, \bibinfo {author} {\bibfnamefont {R.}~\bibnamefont {Gold}}, \ and\
  \bibinfo {author} {\bibfnamefont {A.}~\bibnamefont {Eichhorn}},\ }\href
  {\doibase 10.1088/1475-7516/2019/06/029} {\bibfield  {journal} {\bibinfo
  {journal} {JCAP}\ }\textbf {\bibinfo {volume} {06}},\ \bibinfo {pages} {029}
  (\bibinfo {year} {2019})},\ \Eprint {http://arxiv.org/abs/1904.07133}
  {arXiv:1904.07133 [gr-qc]} \BibitemShut {NoStop}%
\bibitem [{\citenamefont {Banados}\ \emph {et~al.}(1992)\citenamefont
  {Banados}, \citenamefont {Teitelboim},\ and\ \citenamefont
  {Zanelli}}]{Banados:1992wn}%
  \BibitemOpen
  \bibfield  {author} {\bibinfo {author} {\bibfnamefont {M.}~\bibnamefont
  {Banados}}, \bibinfo {author} {\bibfnamefont {C.}~\bibnamefont {Teitelboim}},
  \ and\ \bibinfo {author} {\bibfnamefont {J.}~\bibnamefont {Zanelli}},\ }\href
  {\doibase 10.1103/PhysRevLett.69.1849} {\bibfield  {journal} {\bibinfo
  {journal} {Phys. Rev. Lett.}\ }\textbf {\bibinfo {volume} {69}},\ \bibinfo
  {pages} {1849} (\bibinfo {year} {1992})},\ \Eprint
  {http://arxiv.org/abs/hep-th/9204099} {arXiv:hep-th/9204099} \BibitemShut
  {NoStop}%
\bibitem [{\citenamefont {Konoplya}\ and\ \citenamefont
  {Zhidenko}(2020)}]{Konoplya:2020ibi}%
  \BibitemOpen
  \bibfield  {author} {\bibinfo {author} {\bibfnamefont {R.~A.}\ \bibnamefont
  {Konoplya}}\ and\ \bibinfo {author} {\bibfnamefont {A.}~\bibnamefont
  {Zhidenko}},\ }\href {\doibase 10.1103/PhysRevD.102.064004} {\bibfield
  {journal} {\bibinfo  {journal} {Phys. Rev. D}\ }\textbf {\bibinfo {volume}
  {102}},\ \bibinfo {pages} {064004} (\bibinfo {year} {2020})},\ \Eprint
  {http://arxiv.org/abs/2003.12171} {arXiv:2003.12171 [gr-qc]} \BibitemShut
  {NoStop}%
\bibitem [{\citenamefont {Skvortsova}(2023)}]{Skvortsova:2023zmj}%
  \BibitemOpen
  \bibfield  {author} {\bibinfo {author} {\bibfnamefont {M.}~\bibnamefont
  {Skvortsova}},\ }\href@noop {} {\  (\bibinfo {year} {2023})},\ \Eprint
  {http://arxiv.org/abs/2311.11650} {arXiv:2311.11650 [gr-qc]} \BibitemShut
  {NoStop}%
\bibitem [{\citenamefont {Cardoso}\ and\ \citenamefont
  {Lemos}(2003)}]{Cardoso:2003sw}%
  \BibitemOpen
  \bibfield  {author} {\bibinfo {author} {\bibfnamefont {V.}~\bibnamefont
  {Cardoso}}\ and\ \bibinfo {author} {\bibfnamefont {J.~P.~S.}\ \bibnamefont
  {Lemos}},\ }\href {\doibase 10.1103/PhysRevD.67.084020} {\bibfield  {journal}
  {\bibinfo  {journal} {Phys. Rev. D}\ }\textbf {\bibinfo {volume} {67}},\
  \bibinfo {pages} {084020} (\bibinfo {year} {2003})},\ \Eprint
  {http://arxiv.org/abs/gr-qc/0301078} {arXiv:gr-qc/0301078} \BibitemShut
  {NoStop}%
\bibitem [{\citenamefont {Molina}(2003)}]{Molina:2003ff}%
  \BibitemOpen
  \bibfield  {author} {\bibinfo {author} {\bibfnamefont {C.}~\bibnamefont
  {Molina}},\ }\href {\doibase 10.1103/PhysRevD.68.064007} {\bibfield
  {journal} {\bibinfo  {journal} {Phys. Rev. D}\ }\textbf {\bibinfo {volume}
  {68}},\ \bibinfo {pages} {064007} (\bibinfo {year} {2003})},\ \Eprint
  {http://arxiv.org/abs/gr-qc/0304053} {arXiv:gr-qc/0304053} \BibitemShut
  {NoStop}%
\bibitem [{\citenamefont {Konoplya}\ and\ \citenamefont
  {Zhidenko}(2023)}]{Konoplya:2023moy}%
  \BibitemOpen
  \bibfield  {author} {\bibinfo {author} {\bibfnamefont {R.~A.}\ \bibnamefont
  {Konoplya}}\ and\ \bibinfo {author} {\bibfnamefont {A.}~\bibnamefont
  {Zhidenko}},\ }\href {\doibase 10.1088/1361-6382/ad0a52} {\bibfield
  {journal} {\bibinfo  {journal} {Class. Quant. Grav.}\ }\textbf {\bibinfo
  {volume} {40}},\ \bibinfo {pages} {245005} (\bibinfo {year} {2023})},\
  \Eprint {http://arxiv.org/abs/2309.02560} {arXiv:2309.02560 [gr-qc]}
  \BibitemShut {NoStop}%
\bibitem [{\citenamefont {Malik}(0)}]{doi:10.1142/S0217751X24500246}%
  \BibitemOpen
  \bibfield  {author} {\bibinfo {author} {\bibfnamefont {Z.}~\bibnamefont
  {Malik}},\ }\href {\doibase 10.1142/S0217751X24500246} {\bibfield  {journal}
  {\bibinfo  {journal} {International Journal of Modern Physics A}\ }\textbf
  {\bibinfo {volume} {0}},\ \bibinfo {pages} {2450024} (\bibinfo {year}
  {0})}\BibitemShut {NoStop}%
\bibitem [{\citenamefont {Bolokhov}(2024)}]{Bolokhov:2023bwm}%
  \BibitemOpen
  \bibfield  {author} {\bibinfo {author} {\bibfnamefont {S.~V.}\ \bibnamefont
  {Bolokhov}},\ }\href {\doibase 10.1103/PhysRevD.110.024010} {\bibfield
  {journal} {\bibinfo  {journal} {Phys. Rev. D}\ }\textbf {\bibinfo {volume}
  {110}},\ \bibinfo {pages} {024010} (\bibinfo {year} {2024})},\ \Eprint
  {http://arxiv.org/abs/2311.05503} {arXiv:2311.05503 [gr-qc]} \BibitemShut
  {NoStop}%
\bibitem [{\citenamefont {Cardoso}\ \emph {et~al.}(2009)\citenamefont
  {Cardoso}, \citenamefont {Miranda}, \citenamefont {Berti}, \citenamefont
  {Witek},\ and\ \citenamefont {Zanchin}}]{Cardoso:2008bp}%
  \BibitemOpen
  \bibfield  {author} {\bibinfo {author} {\bibfnamefont {V.}~\bibnamefont
  {Cardoso}}, \bibinfo {author} {\bibfnamefont {A.~S.}\ \bibnamefont
  {Miranda}}, \bibinfo {author} {\bibfnamefont {E.}~\bibnamefont {Berti}},
  \bibinfo {author} {\bibfnamefont {H.}~\bibnamefont {Witek}}, \ and\ \bibinfo
  {author} {\bibfnamefont {V.~T.}\ \bibnamefont {Zanchin}},\ }\href {\doibase
  10.1103/PhysRevD.79.064016} {\bibfield  {journal} {\bibinfo  {journal} {Phys.
  Rev. D}\ }\textbf {\bibinfo {volume} {79}},\ \bibinfo {pages} {064016}
  (\bibinfo {year} {2009})},\ \Eprint {http://arxiv.org/abs/0812.1806}
  {arXiv:0812.1806 [hep-th]} \BibitemShut {NoStop}%
\bibitem [{\citenamefont {Khanna}\ and\ \citenamefont
  {Price}(2017)}]{Khanna:2016yow}%
  \BibitemOpen
  \bibfield  {author} {\bibinfo {author} {\bibfnamefont {G.}~\bibnamefont
  {Khanna}}\ and\ \bibinfo {author} {\bibfnamefont {R.~H.}\ \bibnamefont
  {Price}},\ }\href {\doibase 10.1103/PhysRevD.95.081501} {\bibfield  {journal}
  {\bibinfo  {journal} {Phys. Rev. D}\ }\textbf {\bibinfo {volume} {95}},\
  \bibinfo {pages} {081501} (\bibinfo {year} {2017})},\ \Eprint
  {http://arxiv.org/abs/1609.00083} {arXiv:1609.00083 [gr-qc]} \BibitemShut
  {NoStop}%
\bibitem [{\citenamefont {Konoplya}\ \emph
  {et~al.}(2019{\natexlab{a}})\citenamefont {Konoplya}, \citenamefont
  {Zinhailo},\ and\ \citenamefont {Stuchlík}}]{Konoplya:2019hml}%
  \BibitemOpen
  \bibfield  {author} {\bibinfo {author} {\bibfnamefont {R.~A.}\ \bibnamefont
  {Konoplya}}, \bibinfo {author} {\bibfnamefont {A.~F.}\ \bibnamefont
  {Zinhailo}}, \ and\ \bibinfo {author} {\bibfnamefont {Z.}~\bibnamefont
  {Stuchlík}},\ }\href {\doibase 10.1103/PhysRevD.99.124042} {\bibfield
  {journal} {\bibinfo  {journal} {Phys. Rev. D}\ }\textbf {\bibinfo {volume}
  {99}},\ \bibinfo {pages} {124042} (\bibinfo {year} {2019}{\natexlab{a}})},\
  \Eprint {http://arxiv.org/abs/1903.03483} {arXiv:1903.03483 [gr-qc]}
  \BibitemShut {NoStop}%
\bibitem [{\citenamefont {Bolokhov}(2023)}]{Bolokhov:2023dxq}%
  \BibitemOpen
  \bibfield  {author} {\bibinfo {author} {\bibfnamefont {S.~V.}\ \bibnamefont
  {Bolokhov}},\ }\href@noop {} {\  (\bibinfo {year} {2023})},\ \Eprint
  {http://arxiv.org/abs/2310.12326} {arXiv:2310.12326 [gr-qc]} \BibitemShut
  {NoStop}%
\bibitem [{\citenamefont {Konoplya}\ and\ \citenamefont
  {Stuchlík}(2017)}]{Konoplya:2017wot}%
  \BibitemOpen
  \bibfield  {author} {\bibinfo {author} {\bibfnamefont {R.~A.}\ \bibnamefont
  {Konoplya}}\ and\ \bibinfo {author} {\bibfnamefont {Z.}~\bibnamefont
  {Stuchlík}},\ }\href {\doibase 10.1016/j.physletb.2017.06.015} {\bibfield
  {journal} {\bibinfo  {journal} {Phys. Lett. B}\ }\textbf {\bibinfo {volume}
  {771}},\ \bibinfo {pages} {597} (\bibinfo {year} {2017})},\ \Eprint
  {http://arxiv.org/abs/1705.05928} {arXiv:1705.05928 [gr-qc]} \BibitemShut
  {NoStop}%
\bibitem [{\citenamefont {Konoplya}(2023{\natexlab{b}})}]{Konoplya:2022gjp}%
  \BibitemOpen
  \bibfield  {author} {\bibinfo {author} {\bibfnamefont {R.~A.}\ \bibnamefont
  {Konoplya}},\ }\href {\doibase 10.1016/j.physletb.2023.137674} {\bibfield
  {journal} {\bibinfo  {journal} {Phys. Lett. B}\ }\textbf {\bibinfo {volume}
  {838}},\ \bibinfo {pages} {137674} (\bibinfo {year} {2023}{\natexlab{b}})},\
  \Eprint {http://arxiv.org/abs/2210.08373} {arXiv:2210.08373 [gr-qc]}
  \BibitemShut {NoStop}%
\bibitem [{\citenamefont {Takahashi}\ and\ \citenamefont
  {Soda}(2012)}]{Takahashi:2011du}%
  \BibitemOpen
  \bibfield  {author} {\bibinfo {author} {\bibfnamefont {T.}~\bibnamefont
  {Takahashi}}\ and\ \bibinfo {author} {\bibfnamefont {J.}~\bibnamefont
  {Soda}},\ }\href {\doibase 10.1088/0264-9381/29/3/035008} {\bibfield
  {journal} {\bibinfo  {journal} {Class. Quant. Grav.}\ }\textbf {\bibinfo
  {volume} {29}},\ \bibinfo {pages} {035008} (\bibinfo {year} {2012})},\
  \Eprint {http://arxiv.org/abs/1108.5041} {arXiv:1108.5041 [hep-th]}
  \BibitemShut {NoStop}%
\bibitem [{\citenamefont {Dotti}\ and\ \citenamefont
  {Gleiser}(2005)}]{Dotti:2004sh}%
  \BibitemOpen
  \bibfield  {author} {\bibinfo {author} {\bibfnamefont {G.}~\bibnamefont
  {Dotti}}\ and\ \bibinfo {author} {\bibfnamefont {R.~J.}\ \bibnamefont
  {Gleiser}},\ }\href {\doibase 10.1088/0264-9381/22/1/L01} {\bibfield
  {journal} {\bibinfo  {journal} {Class. Quant. Grav.}\ }\textbf {\bibinfo
  {volume} {22}},\ \bibinfo {pages} {L1} (\bibinfo {year} {2005})},\ \Eprint
  {http://arxiv.org/abs/gr-qc/0409005} {arXiv:gr-qc/0409005} \BibitemShut
  {NoStop}%
\bibitem [{\citenamefont {Gleiser}\ and\ \citenamefont
  {Dotti}(2005)}]{Gleiser:2005ra}%
  \BibitemOpen
  \bibfield  {author} {\bibinfo {author} {\bibfnamefont {R.~J.}\ \bibnamefont
  {Gleiser}}\ and\ \bibinfo {author} {\bibfnamefont {G.}~\bibnamefont
  {Dotti}},\ }\href {\doibase 10.1103/PhysRevD.72.124002} {\bibfield  {journal}
  {\bibinfo  {journal} {Phys. Rev. D}\ }\textbf {\bibinfo {volume} {72}},\
  \bibinfo {pages} {124002} (\bibinfo {year} {2005})},\ \Eprint
  {http://arxiv.org/abs/gr-qc/0510069} {arXiv:gr-qc/0510069} \BibitemShut
  {NoStop}%
\bibitem [{\citenamefont {Cuyubamba}\ \emph {et~al.}(2016)\citenamefont
  {Cuyubamba}, \citenamefont {Konoplya},\ and\ \citenamefont
  {Zhidenko}}]{Cuyubamba:2016cug}%
  \BibitemOpen
  \bibfield  {author} {\bibinfo {author} {\bibfnamefont {M.~A.}\ \bibnamefont
  {Cuyubamba}}, \bibinfo {author} {\bibfnamefont {R.~A.}\ \bibnamefont
  {Konoplya}}, \ and\ \bibinfo {author} {\bibfnamefont {A.}~\bibnamefont
  {Zhidenko}},\ }\href {\doibase 10.1103/PhysRevD.93.104053} {\bibfield
  {journal} {\bibinfo  {journal} {Phys. Rev. D}\ }\textbf {\bibinfo {volume}
  {93}},\ \bibinfo {pages} {104053} (\bibinfo {year} {2016})},\ \Eprint
  {http://arxiv.org/abs/1604.03604} {arXiv:1604.03604 [gr-qc]} \BibitemShut
  {NoStop}%
\bibitem [{\citenamefont {Konoplya}\ and\ \citenamefont
  {Zhidenko}(2017)}]{Konoplya:2017lhs}%
  \BibitemOpen
  \bibfield  {author} {\bibinfo {author} {\bibfnamefont {R.~A.}\ \bibnamefont
  {Konoplya}}\ and\ \bibinfo {author} {\bibfnamefont {A.}~\bibnamefont
  {Zhidenko}},\ }\href {\doibase 10.1088/1475-7516/2017/05/050} {\bibfield
  {journal} {\bibinfo  {journal} {JCAP}\ }\textbf {\bibinfo {volume} {05}},\
  \bibinfo {pages} {050} (\bibinfo {year} {2017})},\ \Eprint
  {http://arxiv.org/abs/1705.01656} {arXiv:1705.01656 [hep-th]} \BibitemShut
  {NoStop}%
\bibitem [{\citenamefont {Konoplya}\ and\ \citenamefont
  {Zhidenko}(2008)}]{Konoplya:2008ix}%
  \BibitemOpen
  \bibfield  {author} {\bibinfo {author} {\bibfnamefont {R.~A.}\ \bibnamefont
  {Konoplya}}\ and\ \bibinfo {author} {\bibfnamefont {A.}~\bibnamefont
  {Zhidenko}},\ }\href {\doibase 10.1103/PhysRevD.77.104004} {\bibfield
  {journal} {\bibinfo  {journal} {Phys. Rev. D}\ }\textbf {\bibinfo {volume}
  {77}},\ \bibinfo {pages} {104004} (\bibinfo {year} {2008})},\ \Eprint
  {http://arxiv.org/abs/0802.0267} {arXiv:0802.0267 [hep-th]} \BibitemShut
  {NoStop}%
\bibitem [{\citenamefont {Paul}(2024)}]{Paul:2023eep}%
  \BibitemOpen
  \bibfield  {author} {\bibinfo {author} {\bibfnamefont {P.}~\bibnamefont
  {Paul}},\ }\href {\doibase 10.1140/epjc/s10052-024-12563-6} {\bibfield
  {journal} {\bibinfo  {journal} {Eur. Phys. J. C}\ }\textbf {\bibinfo {volume}
  {84}},\ \bibinfo {pages} {218} (\bibinfo {year} {2024})},\ \Eprint
  {http://arxiv.org/abs/2312.16479} {arXiv:2312.16479 [gr-qc]} \BibitemShut
  {NoStop}%
\bibitem [{\citenamefont {Konoplya}\ and\ \citenamefont
  {Zinhailo}(2020)}]{Konoplya:2020bxa}%
  \BibitemOpen
  \bibfield  {author} {\bibinfo {author} {\bibfnamefont {R.~A.}\ \bibnamefont
  {Konoplya}}\ and\ \bibinfo {author} {\bibfnamefont {A.~F.}\ \bibnamefont
  {Zinhailo}},\ }\href {\doibase 10.1140/epjc/s10052-020-08639-8} {\bibfield
  {journal} {\bibinfo  {journal} {Eur. Phys. J. C}\ }\textbf {\bibinfo {volume}
  {80}},\ \bibinfo {pages} {1049} (\bibinfo {year} {2020})},\ \Eprint
  {http://arxiv.org/abs/2003.01188} {arXiv:2003.01188 [gr-qc]} \BibitemShut
  {NoStop}%
\bibitem [{\citenamefont {Zhidenko}(2004)}]{Zhidenko:2003wq}%
  \BibitemOpen
  \bibfield  {author} {\bibinfo {author} {\bibfnamefont {A.}~\bibnamefont
  {Zhidenko}},\ }\href {\doibase 10.1088/0264-9381/21/1/019} {\bibfield
  {journal} {\bibinfo  {journal} {Class. Quant. Grav.}\ }\textbf {\bibinfo
  {volume} {21}},\ \bibinfo {pages} {273} (\bibinfo {year} {2004})},\ \Eprint
  {http://arxiv.org/abs/gr-qc/0307012} {arXiv:gr-qc/0307012} \BibitemShut
  {NoStop}%
\bibitem [{\citenamefont {Chen}\ \emph {et~al.}(2023)\citenamefont {Chen},
  \citenamefont {Chen}, \citenamefont {Ho},\ and\ \citenamefont
  {Tseng}}]{Chen:2022nlw}%
  \BibitemOpen
  \bibfield  {author} {\bibinfo {author} {\bibfnamefont {C.-Y.}\ \bibnamefont
  {Chen}}, \bibinfo {author} {\bibfnamefont {Y.-J.}\ \bibnamefont {Chen}},
  \bibinfo {author} {\bibfnamefont {M.-Y.}\ \bibnamefont {Ho}}, \ and\ \bibinfo
  {author} {\bibfnamefont {Y.-H.}\ \bibnamefont {Tseng}},\ }\href {\doibase
  10.1016/j.physletb.2023.138153} {\bibfield  {journal} {\bibinfo  {journal}
  {Phys. Lett. B}\ }\textbf {\bibinfo {volume} {845}},\ \bibinfo {pages}
  {138153} (\bibinfo {year} {2023})},\ \Eprint
  {http://arxiv.org/abs/2212.10028} {arXiv:2212.10028 [gr-qc]} \BibitemShut
  {NoStop}%
\bibitem [{\citenamefont {Silva}\ and\ \citenamefont
  {Glampedakis}(2020)}]{Silva:2019scu}%
  \BibitemOpen
  \bibfield  {author} {\bibinfo {author} {\bibfnamefont {H.~O.}\ \bibnamefont
  {Silva}}\ and\ \bibinfo {author} {\bibfnamefont {K.}~\bibnamefont
  {Glampedakis}},\ }\href {\doibase 10.1103/PhysRevD.101.044051} {\bibfield
  {journal} {\bibinfo  {journal} {Phys. Rev. D}\ }\textbf {\bibinfo {volume}
  {101}},\ \bibinfo {pages} {044051} (\bibinfo {year} {2020})},\ \Eprint
  {http://arxiv.org/abs/1912.09286} {arXiv:1912.09286 [gr-qc]} \BibitemShut
  {NoStop}%
\bibitem [{\citenamefont {Toshmatov}\ \emph {et~al.}(2019)\citenamefont
  {Toshmatov}, \citenamefont {Stuchlík}, \citenamefont {Ahmedov},\ and\
  \citenamefont {Malafarina}}]{Toshmatov:2019gxg}%
  \BibitemOpen
  \bibfield  {author} {\bibinfo {author} {\bibfnamefont {B.}~\bibnamefont
  {Toshmatov}}, \bibinfo {author} {\bibfnamefont {Z.}~\bibnamefont
  {Stuchlík}}, \bibinfo {author} {\bibfnamefont {B.}~\bibnamefont {Ahmedov}},
  \ and\ \bibinfo {author} {\bibfnamefont {D.}~\bibnamefont {Malafarina}},\
  }\href {\doibase 10.1103/PhysRevD.99.064043} {\bibfield  {journal} {\bibinfo
  {journal} {Phys. Rev. D}\ }\textbf {\bibinfo {volume} {99}},\ \bibinfo
  {pages} {064043} (\bibinfo {year} {2019})},\ \Eprint
  {http://arxiv.org/abs/1903.03778} {arXiv:1903.03778 [gr-qc]} \BibitemShut
  {NoStop}%
\bibitem [{\citenamefont {Allahyari}\ \emph {et~al.}(2019)\citenamefont
  {Allahyari}, \citenamefont {Firouzjahi},\ and\ \citenamefont
  {Mashhoon}}]{Allahyari:2018cmg}%
  \BibitemOpen
  \bibfield  {author} {\bibinfo {author} {\bibfnamefont {A.}~\bibnamefont
  {Allahyari}}, \bibinfo {author} {\bibfnamefont {H.}~\bibnamefont
  {Firouzjahi}}, \ and\ \bibinfo {author} {\bibfnamefont {B.}~\bibnamefont
  {Mashhoon}},\ }\href {\doibase 10.1103/PhysRevD.99.044005} {\bibfield
  {journal} {\bibinfo  {journal} {Phys. Rev. D}\ }\textbf {\bibinfo {volume}
  {99}},\ \bibinfo {pages} {044005} (\bibinfo {year} {2019})},\ \Eprint
  {http://arxiv.org/abs/1812.03376} {arXiv:1812.03376 [gr-qc]} \BibitemShut
  {NoStop}%
\bibitem [{\citenamefont {Berglund}\ \emph {et~al.}(2012)\citenamefont
  {Berglund}, \citenamefont {Bhattacharyya},\ and\ \citenamefont
  {Mattingly}}]{Berglund:2012bu}%
  \BibitemOpen
  \bibfield  {author} {\bibinfo {author} {\bibfnamefont {P.}~\bibnamefont
  {Berglund}}, \bibinfo {author} {\bibfnamefont {J.}~\bibnamefont
  {Bhattacharyya}}, \ and\ \bibinfo {author} {\bibfnamefont {D.}~\bibnamefont
  {Mattingly}},\ }\href {\doibase 10.1103/PhysRevD.85.124019} {\bibfield
  {journal} {\bibinfo  {journal} {Phys. Rev. D}\ }\textbf {\bibinfo {volume}
  {85}},\ \bibinfo {pages} {124019} (\bibinfo {year} {2012})},\ \Eprint
  {http://arxiv.org/abs/1202.4497} {arXiv:1202.4497 [hep-th]} \BibitemShut
  {NoStop}%
\bibitem [{\citenamefont {Kokkotas}\ and\ \citenamefont
  {Schmidt}(1999)}]{Kokkotas:1999bd}%
  \BibitemOpen
  \bibfield  {author} {\bibinfo {author} {\bibfnamefont {K.~D.}\ \bibnamefont
  {Kokkotas}}\ and\ \bibinfo {author} {\bibfnamefont {B.~G.}\ \bibnamefont
  {Schmidt}},\ }\href {\doibase 10.12942/lrr-1999-2} {\bibfield  {journal}
  {\bibinfo  {journal} {Living Rev. Rel.}\ }\textbf {\bibinfo {volume} {2}},\
  \bibinfo {pages} {2} (\bibinfo {year} {1999})},\ \Eprint
  {http://arxiv.org/abs/gr-qc/9909058} {arXiv:gr-qc/9909058} \BibitemShut
  {NoStop}%
\bibitem [{\citenamefont {Konoplya}\ and\ \citenamefont
  {Zhidenko}(2011)}]{Konoplya:2011qq}%
  \BibitemOpen
  \bibfield  {author} {\bibinfo {author} {\bibfnamefont {R.~A.}\ \bibnamefont
  {Konoplya}}\ and\ \bibinfo {author} {\bibfnamefont {A.}~\bibnamefont
  {Zhidenko}},\ }\href {\doibase 10.1103/RevModPhys.83.793} {\bibfield
  {journal} {\bibinfo  {journal} {Rev. Mod. Phys.}\ }\textbf {\bibinfo {volume}
  {83}},\ \bibinfo {pages} {793} (\bibinfo {year} {2011})},\ \Eprint
  {http://arxiv.org/abs/1102.4014} {arXiv:1102.4014 [gr-qc]} \BibitemShut
  {NoStop}%
\bibitem [{\citenamefont {Schutz}\ and\ \citenamefont
  {Will}(1985)}]{Schutz:1985km}%
  \BibitemOpen
  \bibfield  {author} {\bibinfo {author} {\bibfnamefont {B.~F.}\ \bibnamefont
  {Schutz}}\ and\ \bibinfo {author} {\bibfnamefont {C.~M.}\ \bibnamefont
  {Will}},\ }\href {\doibase 10.1086/184453} {\bibfield  {journal} {\bibinfo
  {journal} {Astrophys. J. Lett.}\ }\textbf {\bibinfo {volume} {291}},\
  \bibinfo {pages} {L33} (\bibinfo {year} {1985})}\BibitemShut {NoStop}%
\bibitem [{\citenamefont {Konoplya}\ \emph
  {et~al.}(2019{\natexlab{b}})\citenamefont {Konoplya}, \citenamefont
  {Zhidenko},\ and\ \citenamefont {Zinhailo}}]{Konoplya:2019hlu}%
  \BibitemOpen
  \bibfield  {author} {\bibinfo {author} {\bibfnamefont {R.~A.}\ \bibnamefont
  {Konoplya}}, \bibinfo {author} {\bibfnamefont {A.}~\bibnamefont {Zhidenko}},
  \ and\ \bibinfo {author} {\bibfnamefont {A.~F.}\ \bibnamefont {Zinhailo}},\
  }\href {\doibase 10.1088/1361-6382/ab2e25} {\bibfield  {journal} {\bibinfo
  {journal} {Class. Quant. Grav.}\ }\textbf {\bibinfo {volume} {36}},\ \bibinfo
  {pages} {155002} (\bibinfo {year} {2019}{\natexlab{b}})},\ \Eprint
  {http://arxiv.org/abs/1904.10333} {arXiv:1904.10333 [gr-qc]} \BibitemShut
  {NoStop}%
\bibitem [{\citenamefont {Iyer}\ and\ \citenamefont
  {Will}(1987)}]{Iyer:1986np}%
  \BibitemOpen
  \bibfield  {author} {\bibinfo {author} {\bibfnamefont {S.}~\bibnamefont
  {Iyer}}\ and\ \bibinfo {author} {\bibfnamefont {C.~M.}\ \bibnamefont
  {Will}},\ }\href {\doibase 10.1103/PhysRevD.35.3621} {\bibfield  {journal}
  {\bibinfo  {journal} {Phys. Rev. D}\ }\textbf {\bibinfo {volume} {35}},\
  \bibinfo {pages} {3621} (\bibinfo {year} {1987})}\BibitemShut {NoStop}%
\bibitem [{\citenamefont {Konoplya}(2003)}]{Konoplya:2003ii}%
  \BibitemOpen
  \bibfield  {author} {\bibinfo {author} {\bibfnamefont {R.~A.}\ \bibnamefont
  {Konoplya}},\ }\href {\doibase 10.1103/PhysRevD.68.024018} {\bibfield
  {journal} {\bibinfo  {journal} {Phys. Rev. D}\ }\textbf {\bibinfo {volume}
  {68}},\ \bibinfo {pages} {024018} (\bibinfo {year} {2003})},\ \Eprint
  {http://arxiv.org/abs/gr-qc/0303052} {arXiv:gr-qc/0303052} \BibitemShut
  {NoStop}%
\bibitem [{\citenamefont {Matyjasek}\ and\ \citenamefont
  {Opala}(2017)}]{Matyjasek:2017psv}%
  \BibitemOpen
  \bibfield  {author} {\bibinfo {author} {\bibfnamefont {J.}~\bibnamefont
  {Matyjasek}}\ and\ \bibinfo {author} {\bibfnamefont {M.}~\bibnamefont
  {Opala}},\ }\href {\doibase 10.1103/PhysRevD.96.024011} {\bibfield  {journal}
  {\bibinfo  {journal} {Phys. Rev. D}\ }\textbf {\bibinfo {volume} {96}},\
  \bibinfo {pages} {024011} (\bibinfo {year} {2017})},\ \Eprint
  {http://arxiv.org/abs/1704.00361} {arXiv:1704.00361 [gr-qc]} \BibitemShut
  {NoStop}%
\bibitem [{\citenamefont {Konoplya}(2002)}]{Konoplya:2001ji}%
  \BibitemOpen
  \bibfield  {author} {\bibinfo {author} {\bibfnamefont {R.~A.}\ \bibnamefont
  {Konoplya}},\ }\href {\doibase 10.1023/A:1015347628961} {\bibfield  {journal}
  {\bibinfo  {journal} {Gen. Rel. Grav.}\ }\textbf {\bibinfo {volume} {34}},\
  \bibinfo {pages} {329} (\bibinfo {year} {2002})},\ \Eprint
  {http://arxiv.org/abs/gr-qc/0109096} {arXiv:gr-qc/0109096} \BibitemShut
  {NoStop}%
\bibitem [{\citenamefont {Konoplya}\ and\ \citenamefont
  {Zhidenko}(2007)}]{Konoplya:2006ar}%
  \BibitemOpen
  \bibfield  {author} {\bibinfo {author} {\bibfnamefont {R.~A.}\ \bibnamefont
  {Konoplya}}\ and\ \bibinfo {author} {\bibfnamefont {A.}~\bibnamefont
  {Zhidenko}},\ }\href {\doibase 10.1016/j.physletb.2007.03.018} {\bibfield
  {journal} {\bibinfo  {journal} {Phys. Lett. B}\ }\textbf {\bibinfo {volume}
  {648}},\ \bibinfo {pages} {236} (\bibinfo {year} {2007})},\ \Eprint
  {http://arxiv.org/abs/hep-th/0611226} {arXiv:hep-th/0611226} \BibitemShut
  {NoStop}%
\bibitem [{\citenamefont {Konoplya}(2018)}]{Konoplya:2018ala}%
  \BibitemOpen
  \bibfield  {author} {\bibinfo {author} {\bibfnamefont {R.~A.}\ \bibnamefont
  {Konoplya}},\ }\href {\doibase 10.1016/j.physletb.2018.07.025} {\bibfield
  {journal} {\bibinfo  {journal} {Phys. Lett. B}\ }\textbf {\bibinfo {volume}
  {784}},\ \bibinfo {pages} {43} (\bibinfo {year} {2018})},\ \Eprint
  {http://arxiv.org/abs/1805.04718} {arXiv:1805.04718 [gr-qc]} \BibitemShut
  {NoStop}%
\bibitem [{\citenamefont {Kodama}\ \emph {et~al.}(2010)\citenamefont {Kodama},
  \citenamefont {Konoplya},\ and\ \citenamefont {Zhidenko}}]{Kodama:2009bf}%
  \BibitemOpen
  \bibfield  {author} {\bibinfo {author} {\bibfnamefont {H.}~\bibnamefont
  {Kodama}}, \bibinfo {author} {\bibfnamefont {R.~A.}\ \bibnamefont
  {Konoplya}}, \ and\ \bibinfo {author} {\bibfnamefont {A.}~\bibnamefont
  {Zhidenko}},\ }\href {\doibase 10.1103/PhysRevD.81.044007} {\bibfield
  {journal} {\bibinfo  {journal} {Phys. Rev. D}\ }\textbf {\bibinfo {volume}
  {81}},\ \bibinfo {pages} {044007} (\bibinfo {year} {2010})},\ \Eprint
  {http://arxiv.org/abs/0904.2154} {arXiv:0904.2154 [gr-qc]} \BibitemShut
  {NoStop}%
\bibitem [{\citenamefont {Gong}\ \emph {et~al.}(2023)\citenamefont {Gong},
  \citenamefont {Li}, \citenamefont {Zhang}, \citenamefont {Fu},\ and\
  \citenamefont {Wu}}]{Gong:2023ghh}%
  \BibitemOpen
  \bibfield  {author} {\bibinfo {author} {\bibfnamefont {H.}~\bibnamefont
  {Gong}}, \bibinfo {author} {\bibfnamefont {S.}~\bibnamefont {Li}}, \bibinfo
  {author} {\bibfnamefont {D.}~\bibnamefont {Zhang}}, \bibinfo {author}
  {\bibfnamefont {G.}~\bibnamefont {Fu}}, \ and\ \bibinfo {author}
  {\bibfnamefont {J.-P.}\ \bibnamefont {Wu}},\ }\href@noop {} {\  (\bibinfo
  {year} {2023})},\ \Eprint {http://arxiv.org/abs/2312.17639} {arXiv:2312.17639
  [gr-qc]} \BibitemShut {NoStop}%
\bibitem [{\citenamefont {Fernando}\ and\ \citenamefont
  {Correa}(2012)}]{Fernando:2012yw}%
  \BibitemOpen
  \bibfield  {author} {\bibinfo {author} {\bibfnamefont {S.}~\bibnamefont
  {Fernando}}\ and\ \bibinfo {author} {\bibfnamefont {J.}~\bibnamefont
  {Correa}},\ }\href {\doibase 10.1103/PhysRevD.86.064039} {\bibfield
  {journal} {\bibinfo  {journal} {Phys. Rev. D}\ }\textbf {\bibinfo {volume}
  {86}},\ \bibinfo {pages} {064039} (\bibinfo {year} {2012})},\ \Eprint
  {http://arxiv.org/abs/1208.5442} {arXiv:1208.5442 [gr-qc]} \BibitemShut
  {NoStop}%
\bibitem [{\citenamefont {Dubinsky}\ and\ \citenamefont
  {Zinhailo}(2024)}]{Dubinsky:2024hmn}%
  \BibitemOpen
  \bibfield  {author} {\bibinfo {author} {\bibfnamefont {A.}~\bibnamefont
  {Dubinsky}}\ and\ \bibinfo {author} {\bibfnamefont {A.}~\bibnamefont
  {Zinhailo}},\ }\href@noop {} {\  (\bibinfo {year} {2024})},\ \Eprint
  {http://arxiv.org/abs/2404.01834} {arXiv:2404.01834 [gr-qc]} \BibitemShut
  {NoStop}%
\bibitem [{\citenamefont {Konoplya}\ \emph {et~al.}(2020)\citenamefont
  {Konoplya}, \citenamefont {Zinhailo},\ and\ \citenamefont
  {Stuchlik}}]{Konoplya:2020jgt}%
  \BibitemOpen
  \bibfield  {author} {\bibinfo {author} {\bibfnamefont {R.~A.}\ \bibnamefont
  {Konoplya}}, \bibinfo {author} {\bibfnamefont {A.~F.}\ \bibnamefont
  {Zinhailo}}, \ and\ \bibinfo {author} {\bibfnamefont {Z.}~\bibnamefont
  {Stuchlik}},\ }\href {\doibase 10.1103/PhysRevD.102.044023} {\bibfield
  {journal} {\bibinfo  {journal} {Phys. Rev. D}\ }\textbf {\bibinfo {volume}
  {102}},\ \bibinfo {pages} {044023} (\bibinfo {year} {2020})},\ \Eprint
  {http://arxiv.org/abs/2006.10462} {arXiv:2006.10462 [gr-qc]} \BibitemShut
  {NoStop}%
\bibitem [{\citenamefont {Malik}(2024{\natexlab{a}})}]{Malik:2024voy}%
  \BibitemOpen
  \bibfield  {author} {\bibinfo {author} {\bibfnamefont {Z.}~\bibnamefont
  {Malik}},\ }\href {\doibase 10.1142/S0217751X24500246} {\bibfield  {journal}
  {\bibinfo  {journal} {Int. J. Mod. Phys. A}\ }\textbf {\bibinfo {volume}
  {39}},\ \bibinfo {pages} {2450024} (\bibinfo {year}
  {2024}{\natexlab{a}})}\BibitemShut {NoStop}%
\bibitem [{\citenamefont {Malik}(2024{\natexlab{b}})}]{Malik:2024sxv}%
  \BibitemOpen
  \bibfield  {author} {\bibinfo {author} {\bibfnamefont {Z.}~\bibnamefont
  {Malik}},\ }\href {\doibase 10.1007/s10773-024-05660-5} {\bibfield  {journal}
  {\bibinfo  {journal} {Int. J. Theor. Phys.}\ }\textbf {\bibinfo {volume}
  {63}},\ \bibinfo {pages} {128} (\bibinfo {year}
  {2024}{\natexlab{b}})}\BibitemShut {NoStop}%
\bibitem [{\citenamefont {Gundlach}\ \emph {et~al.}(1994)\citenamefont
  {Gundlach}, \citenamefont {Price},\ and\ \citenamefont
  {Pullin}}]{Gundlach:1993tp}%
  \BibitemOpen
  \bibfield  {author} {\bibinfo {author} {\bibfnamefont {C.}~\bibnamefont
  {Gundlach}}, \bibinfo {author} {\bibfnamefont {R.~H.}\ \bibnamefont {Price}},
  \ and\ \bibinfo {author} {\bibfnamefont {J.}~\bibnamefont {Pullin}},\ }\href
  {\doibase 10.1103/PhysRevD.49.883} {\bibfield  {journal} {\bibinfo  {journal}
  {Phys. Rev. D}\ }\textbf {\bibinfo {volume} {49}},\ \bibinfo {pages} {883}
  (\bibinfo {year} {1994})},\ \Eprint {http://arxiv.org/abs/gr-qc/9307009}
  {arXiv:gr-qc/9307009} \BibitemShut {NoStop}%
\bibitem [{\citenamefont {Konoplya}\ and\ \citenamefont
  {Zhidenko}(2014)}]{Konoplya:2014lha}%
  \BibitemOpen
  \bibfield  {author} {\bibinfo {author} {\bibfnamefont {R.~A.}\ \bibnamefont
  {Konoplya}}\ and\ \bibinfo {author} {\bibfnamefont {A.}~\bibnamefont
  {Zhidenko}},\ }\href {\doibase 10.1103/PhysRevD.90.064048} {\bibfield
  {journal} {\bibinfo  {journal} {Phys. Rev. D}\ }\textbf {\bibinfo {volume}
  {90}},\ \bibinfo {pages} {064048} (\bibinfo {year} {2014})},\ \Eprint
  {http://arxiv.org/abs/1406.0019} {arXiv:1406.0019 [hep-th]} \BibitemShut
  {NoStop}%
\bibitem [{\citenamefont {Ishihara}\ \emph {et~al.}(2008)\citenamefont
  {Ishihara}, \citenamefont {Kimura}, \citenamefont {Konoplya}, \citenamefont
  {Murata}, \citenamefont {Soda},\ and\ \citenamefont
  {Zhidenko}}]{Ishihara:2008re}%
  \BibitemOpen
  \bibfield  {author} {\bibinfo {author} {\bibfnamefont {H.}~\bibnamefont
  {Ishihara}}, \bibinfo {author} {\bibfnamefont {M.}~\bibnamefont {Kimura}},
  \bibinfo {author} {\bibfnamefont {R.~A.}\ \bibnamefont {Konoplya}}, \bibinfo
  {author} {\bibfnamefont {K.}~\bibnamefont {Murata}}, \bibinfo {author}
  {\bibfnamefont {J.}~\bibnamefont {Soda}}, \ and\ \bibinfo {author}
  {\bibfnamefont {A.}~\bibnamefont {Zhidenko}},\ }\href {\doibase
  10.1103/PhysRevD.77.084019} {\bibfield  {journal} {\bibinfo  {journal} {Phys.
  Rev. D}\ }\textbf {\bibinfo {volume} {77}},\ \bibinfo {pages} {084019}
  (\bibinfo {year} {2008})},\ \Eprint {http://arxiv.org/abs/0802.0655}
  {arXiv:0802.0655 [hep-th]} \BibitemShut {NoStop}%
\bibitem [{\citenamefont {Churilova}\ \emph {et~al.}(2021)\citenamefont
  {Churilova}, \citenamefont {Konoplya}, \citenamefont {Stuchlik},\ and\
  \citenamefont {Zhidenko}}]{Churilova:2021tgn}%
  \BibitemOpen
  \bibfield  {author} {\bibinfo {author} {\bibfnamefont {M.~S.}\ \bibnamefont
  {Churilova}}, \bibinfo {author} {\bibfnamefont {R.~A.}\ \bibnamefont
  {Konoplya}}, \bibinfo {author} {\bibfnamefont {Z.}~\bibnamefont {Stuchlik}},
  \ and\ \bibinfo {author} {\bibfnamefont {A.}~\bibnamefont {Zhidenko}},\
  }\href {\doibase 10.1088/1475-7516/2021/10/010} {\bibfield  {journal}
  {\bibinfo  {journal} {JCAP}\ }\textbf {\bibinfo {volume} {10}},\ \bibinfo
  {pages} {010} (\bibinfo {year} {2021})},\ \Eprint
  {http://arxiv.org/abs/2107.05977} {arXiv:2107.05977 [gr-qc]} \BibitemShut
  {NoStop}%
\bibitem [{\citenamefont {Qian}\ \emph {et~al.}(2022)\citenamefont {Qian},
  \citenamefont {Lin}, \citenamefont {Shao}, \citenamefont {Wang},\ and\
  \citenamefont {Yue}}]{Qian:2022kaq}%
  \BibitemOpen
  \bibfield  {author} {\bibinfo {author} {\bibfnamefont {W.-L.}\ \bibnamefont
  {Qian}}, \bibinfo {author} {\bibfnamefont {K.}~\bibnamefont {Lin}}, \bibinfo
  {author} {\bibfnamefont {C.-Y.}\ \bibnamefont {Shao}}, \bibinfo {author}
  {\bibfnamefont {B.}~\bibnamefont {Wang}}, \ and\ \bibinfo {author}
  {\bibfnamefont {R.-H.}\ \bibnamefont {Yue}},\ }\href {\doibase
  10.1140/epjc/s10052-022-10910-z} {\bibfield  {journal} {\bibinfo  {journal}
  {Eur. Phys. J. C}\ }\textbf {\bibinfo {volume} {82}},\ \bibinfo {pages} {931}
  (\bibinfo {year} {2022})},\ \Eprint {http://arxiv.org/abs/2203.04477}
  {arXiv:2203.04477 [gr-qc]} \BibitemShut {NoStop}%
\bibitem [{\citenamefont {Aneesh}\ \emph {et~al.}(2018)\citenamefont {Aneesh},
  \citenamefont {Bose},\ and\ \citenamefont {Kar}}]{Aneesh:2018hlp}%
  \BibitemOpen
  \bibfield  {author} {\bibinfo {author} {\bibfnamefont {S.}~\bibnamefont
  {Aneesh}}, \bibinfo {author} {\bibfnamefont {S.}~\bibnamefont {Bose}}, \ and\
  \bibinfo {author} {\bibfnamefont {S.}~\bibnamefont {Kar}},\ }\href {\doibase
  10.1103/PhysRevD.97.124004} {\bibfield  {journal} {\bibinfo  {journal} {Phys.
  Rev. D}\ }\textbf {\bibinfo {volume} {97}},\ \bibinfo {pages} {124004}
  (\bibinfo {year} {2018})},\ \Eprint {http://arxiv.org/abs/1803.10204}
  {arXiv:1803.10204 [gr-qc]} \BibitemShut {NoStop}%
\bibitem [{\citenamefont {Varghese}\ and\ \citenamefont
  {Kuriakose}(2011)}]{Varghese:2011ku}%
  \BibitemOpen
  \bibfield  {author} {\bibinfo {author} {\bibfnamefont {N.}~\bibnamefont
  {Varghese}}\ and\ \bibinfo {author} {\bibfnamefont {V.~C.}\ \bibnamefont
  {Kuriakose}},\ }\href {\doibase 10.1007/s10714-011-1201-y} {\bibfield
  {journal} {\bibinfo  {journal} {Gen. Rel. Grav.}\ }\textbf {\bibinfo {volume}
  {43}},\ \bibinfo {pages} {2757} (\bibinfo {year} {2011})},\ \Eprint
  {http://arxiv.org/abs/1011.6608} {arXiv:1011.6608 [gr-qc]} \BibitemShut
  {NoStop}%
\bibitem [{\citenamefont {Oshita}(2024)}]{Oshita:2023cjz}%
  \BibitemOpen
  \bibfield  {author} {\bibinfo {author} {\bibfnamefont {N.}~\bibnamefont
  {Oshita}},\ }\href {\doibase 10.1103/PhysRevD.109.104028} {\bibfield
  {journal} {\bibinfo  {journal} {Phys. Rev. D}\ }\textbf {\bibinfo {volume}
  {109}},\ \bibinfo {pages} {104028} (\bibinfo {year} {2024})},\ \Eprint
  {http://arxiv.org/abs/2309.05725} {arXiv:2309.05725 [gr-qc]} \BibitemShut
  {NoStop}%
\bibitem [{\citenamefont {Konoplya}\ and\ \citenamefont
  {Zhidenko}(2024)}]{Konoplya:2024lir}%
  \BibitemOpen
  \bibfield  {author} {\bibinfo {author} {\bibfnamefont {R.~A.}\ \bibnamefont
  {Konoplya}}\ and\ \bibinfo {author} {\bibfnamefont {A.}~\bibnamefont
  {Zhidenko}},\ }\href@noop {} {\  (\bibinfo {year} {2024})},\ \Eprint
  {http://arxiv.org/abs/2406.11694} {arXiv:2406.11694 [gr-qc]} \BibitemShut
  {NoStop}%
\bibitem [{\citenamefont {Rosato}\ \emph {et~al.}(2024)\citenamefont {Rosato},
  \citenamefont {Destounis},\ and\ \citenamefont {Pani}}]{Rosato:2024arw}%
  \BibitemOpen
  \bibfield  {author} {\bibinfo {author} {\bibfnamefont {R.~F.}\ \bibnamefont
  {Rosato}}, \bibinfo {author} {\bibfnamefont {K.}~\bibnamefont {Destounis}}, \
  and\ \bibinfo {author} {\bibfnamefont {P.}~\bibnamefont {Pani}},\ }\href@noop
  {} {\  (\bibinfo {year} {2024})},\ \Eprint {http://arxiv.org/abs/2406.01692}
  {arXiv:2406.01692 [gr-qc]} \BibitemShut {NoStop}%
\end{thebibliography}%
\end{document}